\documentclass[12pt,footnote,twoside]{book}
\usepackage{fancyhdr}
\usepackage{eurosym}
\usepackage[english]{babel}
\usepackage{amsfonts}
\usepackage{amsmath}
\usepackage{amssymb}
\usepackage{graphicx}
\usepackage{subfigure}
\usepackage{txfonts}
\usepackage{makeidx}
\usepackage{array,multirow}

\input{tcilatex}
\begin{document}

\chapter {Magnetic Skyrmions: Theory and Applications}

\centerline{\bf Lalla Btissam Drissi$^{1,2,*}$, El Hassan Saidi$^{1,2}$,
Mosto Bousmina$^{2,3}$, Omar Fassi-Fehri$^{2}$ }

\centerline{\small {$^1$  LPHE, Modeling \& Simulations, Faculty of Science, Mohammed V
University in Rabat, Morocco.}}
\centerline{\small{$^2$ Hassan II Academy of
Science and Technology, Rabat, Morocco.}}
\centerline{\small{$^3$ Euromed Research Institute, Euro-Mediterranean University of Fes,
Fes, Morocco.}}

\centerline{$^*$ \footnotesize \it{Corresponding Author E-mails:
lalla-btissam.drissi@um5.ac.ma, b.drissi@academiesciences.ma}}

\begin{abstract}
Using the field theory method and coherent spin state approach, we
investigate properties of magnetic solitons in spacetime while focussing on
1D kinks, 2D and 3D skyrmions. We also study the case of a rigid skyrmion
dissolved in a magnetic background induced by the electronic spins; and
derive the effective rigid skyrmion equation of motion. We investigate as
well the interaction between an electron and a 3D skyrmion. \newline
\textbf{Keywords}:\emph{\ Geometric phases, magnetic monopoles and topology,
soliton and holonomy, chiral magnets, skyrmion dynamics, numerical and
experimental tests.}

\end{abstract}

\tableofcontents

\section{Introduction}

During the last two decades, the magnetic skyrmions and antiskyrmions have
been subject to an increasing interest in connection with the topological
phase of matter \cite{1,2,3,4}, the spin-tronics \cite{010,011} and quantum
computing \cite{020,021}; as well as in the search for advanced applications
such as racetrack memory, microwave oscillators and logic nanodevices making
skyrmionic states very promising candidates for future low power information
technology devices \cite{022,6,7,8}. Initially proposed by T. Skyrme to
describe hadrons in the theory of quantum chromodynamics \cite{1A},
skyrmions have however been observed in other fields of physics, including
quantum Hall systems \cite{zz001,zz0011}, Bose-Einstein condensates \cite%
{zz002} and liquid crystals \cite{zz003}. In quantum Hall (QH) ferromagnets
for example \cite{2A,3A}, due to the exchange interaction; the electron
spins spontaneously form a fully polarized ferromagnet close to the integer
filling factor $\mathrm{\nu }\simeq 1$; slightly away, other electrons
organize into an intricate spin configuration because of a competitive
interplay between the Coulomb and Zeeman interactions \cite{2A}. Being
quasiparticles, the skyrmions of the QH system condense into a crystalline
form leading to the crystallization of the skyrmions \cite{3AAA1,4A,5A,5AA2}%
; thus opening an important window on promising applications.

In order to overcome the lack of a prototype of a skyrmion-based spintronic
devices for a possible fabrication of nanodevices of data storage and logic
technologies, intense research has been carried out during the last few
years \cite{zz1,zz2}. In this regard, several alternative nano-objects%
\textbf{\ }have been identified to host stable skyrmions at room temperature.%
\textbf{\ }The first experimental observation of crystalline skyrmionic
states was in a three-dimensional metallic ferromagnet\ MnSi with a B20
structure using small angle neutron scattering \cite{1B}. Then, real-space
imaging of the skyrmion has been reported using Lorentz transmission
electron microscopy in non-centrosymmetric magnetic compounds and in thin
films with broken inversion symmetry, including monosilicides,
monogermanides, and their alloys, like Fe$_{1-x}$Co$_{x}$Si \cite{4B}, FeGe
\cite{zz8}, and MnGe \cite{zz9}.\newline

One of the key parameters in the formation of these topologically protected
non-collinear spin textures is the Dzyaloshinskii-Moriya Interaction (DMI)
\cite{zz4,zz5}. Originating from the strong spin-orbit coupling (SOC) at the
interfaces, the DM exchange between atomic spins controls the size and
stability of the induced skyrmions. Depending on the symmetry of the crystal
structures and the skyrmion windings number, the internal spins within a
single skyrmion envelop a sphere in different arrangements \cite{zz6}\textbf{%
.} The in-plane component of the magnetization, in the N\'{e}el skyrmion, is
always pointed in the radial direction \cite{zz11}, while it is oriented
perpendicularly with respect to the position vector in the Bloch skyrmion
\cite{1B}\textbf{. }Different from these two well-known types of skyrmions
are skyrmions with mixed Bloch-N\'{e}el topological spin textures observed
in Co/Pd multilayers \cite{zz12}. Magnetic antiskyrmions, having a more
complex boundary compared to the chiral magnetic boundaries of skyrmions,
exist above room temperature in tetragonal Heusler materials \cite{zz13}.
Higher-order skyrmions should be stabilized in anisotropic frustrated magnet
at zero temperature \cite{zz14} as well as in itinerant magnets with zero
magnetic field \cite{zz15}.

In the quest to miniaturize magnetic storage devices, reduction of
material's dimensions as well as preservation of the stability of magnetic
nano-scale domains are necessary. One possible route to achieve this goal is
the formation of topological protected skyrmions in certain 2D magnetic
materials. To induce magnetic order and tune DMIs in 2D crystal structures,
their centrosymmetric should first be broken using some efficient ways such
a (i) generate one-atom thick hybrids where atoms are mixed in an
alternating manner \cite{zz26,zz27,zz270}, (ii) apply bias voltage or strain
\cite{zz28,zz29,zz31}, (iii) insert adsorbents, impurities and defects \cite%
{zz33,zz35,zz34}. In graphene-like materials, fluorine chemisorption is an
exothermic adsorption that gives rise to stable 2D structures \cite{zz44}
and to long-range magnetism \cite{zz36,zz37}. In semi-fluorinated graphene,
a strong Dzyaloshinskii-Moriya interaction has been predicted with the
presence of ferromagnetic skyrmions \cite{zz46}. The formation of a
nanoskyrmion state in a Sn monolayer on a SiC(0001) surface has been
reported on the basis of a generalized Hubbard model \cite{zz41}. Strong DMI
between the first nearest magnetic germanium neighbors in 2D
semi-fluorinated germanene results in a potential antiferromagnetic skyrmion
\cite{nous21}.

In this bookchapter, we use the coherent spin states\textbf{\ }approach and
the field theory method (continuous limit of lattice magnetic models with
DMI) to revisit some basic aspects and properties of magnetic solitons in
spacetime while focusing on 1d kinks, 2d and 3d spatial
skyrmions/antiskyrmions. We also study the case of a rigid skyrmion
dissolved in a magnetic background induced by the electronic spins of
magnetic atoms like Mn; and derive the effective rigid skyrmion equation of
motion. In this regard, we describe the similarity between, on one hand,
electrons in the electromagnetic background; and, on the other hand, rigid
skyrmions bathing in a texture of magnetic moments. We also investigate the
interaction between electrons and skyrmions as well as the effect of the
spin transfer effect.

This bookchapter is organized as follows: In Section 2, we introduce some
basic tools on quantum SU$\left( 2\right) $ spins and review useful aspects
of their dynamics. In Section 3, we investigate the topological properties
of kinks and 2d space solitons while describing in detail the underside of
the topological structure of these low-dimensional solitons. In Section 4,
we extend the construction to approach topological properties to 3d
skyrmions. In Section 5, we study the dynamics of rigid skyrmions without
and with dissipation; and in Section 6, we use emergent gauge potential
fields to describe the effective dynamics of electrons interacting with the
skyrmion in the presence of a spin transfer torque. We end this study by
making comments and describing perspectives in the study of skyrmions.

\section{Quantum SU$\left( 2\right) $ spin dynamics}

In this section, we review some useful ingredients on the quantum SU$\left(
2\right) $ spin operator, its underlying algebra and its time evolution
while focussing on the interesting spin 1/2 states, concerning electrons in
materials; and on coherent spin states which are at the basis of the study
of skyrmions/antiskyrmions. First, we introduce rapidly the SU$\left(
2\right) $ spin operator $\mathbf{S}$ and the implementation of time
dependence. Then, we investigate the non dissipative dynamics of the spin by
using semi-classical theory approach (coherent states). These tools can be
also viewed as a first step towards the topological study of spin induced
1D, 2D and 3D solitons undertaken in next sections.

\subsection{Quantum spin 1/2 operator and beyond}

We begin by recalling that in non relativistic 3D quantum mechanics, the
spin states $\left \vert S_{z},S\right \rangle $ of spinfull particles are
characterised by two half integers $\left( S_{z},S\right) $, a positive $%
S\geq 0$ and an $S_{z}$ taking $2S+1$ values bounded as $-S\leq S_{z}\leq S$
with integral hoppings. For particles with spin 1/2 like electrons, one
distinguishes two basis vector states $\left \vert \pm \frac{1}{2},\frac{1}{2}%
\right \rangle $ that are eigenvalues of the scaled Pauli matrix $\frac{\hbar
}{2}\sigma _{z}$ and the quadratic (Casimir) operator $\frac{\hbar ^{2}}{4}%
\sum_{a=1}^{3}\sigma _{a}^{2}$, here the three $\frac{\hbar }{2}\sigma _{a}$
with $\sigma _{a}=\vec{\sigma}.\vec{e}_{a}$ are the three components of the
spin 1/2 operator vector\footnote{%
\ For convenience, we often refer to $\vec{\sigma},$ \ $\vec{e}_{i},$ $\vec{%
\sigma}.\vec{e}_{i}=\sigma _{i}$ respectively by bold symbols as $\mathbf{%
\sigma }$, $\mathbf{e}_{i},$ $\mathbf{\sigma }.\mathbf{e}_{i}=\sigma _{i}$.}
$\vec{\sigma}$. From these ingredients, we learn that the average $<S_{z},S|%
\frac{\hbar }{2}\sigma _{z}|S_{z},S>=\hbar S_{z}$ (for short $\left \langle
\frac{\hbar }{2}\sigma _{z}\right \rangle $) is carried by the z-direction
since $S_{z}=\vec{S}.\vec{e}_{z}$ with $\vec{e}_{z}=\left( 0,0,1\right) ^{T}$%
. For generic values of the SU$\left( 2\right) $ spin $S$, the spin operator
reads as $\hbar J_{a}$ where the three $J_{a}$'s are $\left( 2S+1\right)
\times \left( 2S+1\right) $ generators of the SU$\left( 2\right) $ group
satisfying the usual commutation relations $\left[ J_{a},J_{b}\right]
=i\varepsilon _{abc}J^{c}$ with $\varepsilon _{abc}$ standing for the
completely antisymmetric Levi-Civita tensor with non zero value $\varepsilon
_{123}=1$; its inverse is $\varepsilon ^{cba}$ with $\varepsilon ^{123}=-1$.
The time evolution of the spin $\frac{1}{2}$ operator $\hbar \frac{\sigma
_{a}}{2}$ with dynamics governed by a stationary Hamiltonian operator ( $%
dH/dt=0$) is given by the Heisenberg representation of quantum mechanics. In
this non dissipative description, the time dependence of the spin $\frac{1}{2%
}$ operator $\hat{S}_{a}\left( t\right) $ (the hat is to distinguish the
operator $\hat{S}_{a}$ from classical $S_{a}$) is given by
\begin{equation}
\hat{S}_{a}=e^{\frac{i}{\hbar }Ht}\left( \hbar \frac{\sigma _{a}}{2}\right)
e^{-\frac{i}{\hbar }Ht}
\end{equation}%
where the Pauli matrices $\sigma _{a}$ obey the usual commutation relations $%
\left[ \sigma _{a},\sigma _{b}\right] =2i\varepsilon _{abc}\sigma ^{c}$. For
a generic value of the SU$\left( 2\right) $ spin $S$, the above relation
extends as $\hat{S}_{a}=e^{\frac{i}{\hbar }Ht}\left( \hbar J_{a}\right) e^{-%
\frac{i}{\hbar }Ht}$. So, many relations for the spin $1/2$ may be
straightforwardly generalised for generic values $S$ of the SU$\left(
2\right) $ spin. For example, for a spin value $S_{0}$, the $\left(
2S_{0}+1\right) $ states are given by $\left \{ \left \vert
m,S_{0}\right \rangle \right \} $ and are labeled by $-S_{0}\leq m\leq S_{0}$;
one of these states namely $\left \vert S_{0},S_{0}\right \rangle $ is very
special; it is commonly known as the highest weight state (HWS) as it
corresponds to the biggest value $m=S_{0}$; from this state one can generate
all other spin states $\left \vert m,S_{0}\right \rangle $; this feature will
be used when describing coherent spin states. Because of the property $%
\sigma _{a}^{2}=I,$ the square $\hat{S}_{a}^{2}=\frac{\hbar ^{2}}{4}I$ is
time independent; and then the time dynamics of $\hat{S}_{a}\left( t\right) $
is rotational in the sense that $\frac{d\hat{S}_{a}}{dt}$ is given by a
commutator as follows $\frac{d\hat{S}_{a}}{dt}=\frac{i}{\hbar }(H\hat{S}_{a}-%
\hat{S}_{a}H).$ For the example where $H$ is a linearly dependent function
of $\hat{S}_{a}$ like for the Zeeman coupling, the Hamiltonian reads as $%
H_{Z}=\sum_{a}\omega ^{a}\hat{S}_{a}$ (for short $\omega ^{a}\hat{S}_{a}$)
with the $\omega ^{a}$'s are constants referring to the external source%
\footnote{%
For an electron with Zeeman field $B^{a}$, we have $\omega ^{a}=-g\frac{q_{e}%
}{2m_{e}}B^{a}$ with $g=2$ and $q_{e}=-e.$}; then the time evolution of $%
\hat{S}_{a}$ reads, after using the commutation relation $[\hat{S}_{a},\hat{S%
}_{b}]=i\hbar \varepsilon _{abc}\hat{S}^{c}$, as follows%
\begin{equation}
\frac{d\hat{S}_{a}}{dt}=\varepsilon _{abc}\omega ^{b}\hat{S}^{c}\qquad
\Leftrightarrow \qquad \frac{d\mathbf{\hat{S}}}{dt}=\mathbf{\omega }\wedge
\mathbf{\hat{S}}  \label{ms}
\end{equation}%
where appears the Levi-Civita $\varepsilon _{abc}$ which, as we will see
throughout this study, turns out to play an important role in the study of
topological field theory \textrm{\cite{1a,2a}} including solitons and
skyrmions we are interested in here \textrm{\cite{3a,3a1,4a,4a1}}. In this
regards, notice that, along with this $\varepsilon _{abc}$, we will
encounter another completely antisymmetric Levi-Civita tensor namely $%
\epsilon _{\mu _{1}...\mu _{D}}$; it is also due to DM interaction which in
lattice description is given by $(\vec{S}_{\mathbf{r}_{\mu _{2}}}\wedge \vec{%
S}_{\mathbf{r}_{\mu _{1}}}).\vec{d}_{\mu _{3}...\mu _{D-2}}\epsilon ^{\mu
_{1}...\mu _{D}}$; and in continuous limit reads as $\varepsilon
_{abc}S^{b}S_{\mu _{1}\mu _{2}}^{c}d_{\mu _{3}...\mu _{D-2}}^{a}\epsilon
^{\mu _{1}...\mu _{D}}$ where, for convenience, we have set $S_{\mu _{1}\mu
_{2}}^{c}=\mathbf{e}_{\mu _{1}\mu _{2}}.\mathbf{\nabla }S^{c}$ with $\mathbf{%
e}_{\mu _{1}\mu _{2}}=\mathbf{e}_{\mu _{2}}-\mathbf{e}_{\mu _{1}}.$ To
distinguish these two Levi-Civita tensors, we refer to $\varepsilon _{abc}$
as the target space Levi-Civita with SO$\left( 3\right) _{\text{target}}$
symmetry; and to $\epsilon _{\mu _{1}...\mu _{D}}$ as the spacetime
Levi-Civita with SO$\left( 1,D-1\right) $ Lorentz symmetry containing as
subsymmetry the usual space rotation group SO$\left( D-1\right) _{\text{space%
}}$. Notice also that for the case where the Hamiltonian $H(\hat{S})$ is a
general function of the spin, the vector $\omega ^{a}$ is spin dependent and
is given by the gradient $\frac{\partial H}{\partial \hat{S}_{a}}.$

\subsection{Coherent spin states and semi-classical analysis}

\begin{figure}[tbph]
\begin{center}
\includegraphics[width=10cm]{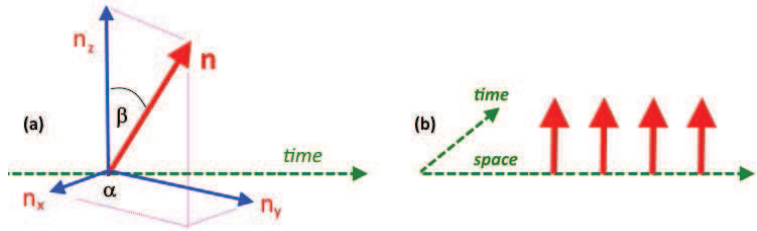}
\end{center}
\par
\vspace{-1cm}
\caption{{\protect \small (a) Components of the spin orientation n; its time
dynamics in presence of a magnetic field is given by Larmor precession. (b)
A configuration of several spins in spacetime.}}
\label{spin1}
\end{figure}

To deal with the semi-classical dynamics of $\mathbf{\hat{S}}\left( t\right)
$ evolved by a Hamiltonian $H(\mathbf{\hat{S})}$, we use the algebra $[\hat{S%
}_{a},\hat{S}_{b}]=i\hbar \varepsilon _{abc}\hat{S}^{c}$ to think of the
quantum spin in terms of a coherent spin state \textrm{\cite{5a}} described
by a (semi) classical vector $\vec{S}=\hbar S\vec{n}$ (no hat) of the
Euclidean $\mathbb{R}^{3};$ see the Figure \textbf{\ref{spin1}}-(a).
This "classical" 3-vector has an amplitude $\hbar S$ and a direction $\vec{n}
$ related to a given unit vector $\vec{n}_{0}$ as $\vec{n}=R\left( \alpha
,\beta ,\gamma \right) \vec{n}_{0};$ and parameterised by $\alpha ,\beta
,\gamma $. In the above relation, the $\vec{n}_{0}$ is thought of as the
north direction of a 2-sphere $\mathbb{S}_{\left( \mathbf{n}\right) }^{2}$
given by the canonical vector $\left( 0,0,1\right) ^{T}$; it is invariant
under the proper rotation; i.e $R_{z}\left( \gamma \right) \vec{n}_{0}=\vec{n%
}_{0}$; and consequently the generic $\vec{n}$ is independent of $\gamma $;
i.e: $\vec{n}=R\left( \alpha ,\beta \right) \vec{n}_{0}.$ Recall that the 3$%
\times $3 matrix $R\left( \alpha ,\beta ,\gamma \right) $ is an SO$\left(
3\right) $ rotation [SO$\left( 3\right) \sim SU\left( 2\right) $] generating
all other points of $\mathbb{S}_{\left( \mathbf{n}\right) }^{2}$
parameterised by $\left( \alpha ,\beta \right) $. In this regards, it is
interesting to recall some useful properties that we list here after as
three points: $\left( 1\right) $ the rotation matrix $R\left( \alpha ,\beta
,\gamma \right) $ can be factorised like $R_{z}\left( \alpha \right)
R_{y}\left( \beta \right) R_{z}\left( \gamma \right) $ where each $%
R_{a}\left( \psi _{a}\right) $ is a rotation $e^{-i\psi _{a}J_{a}}$ around
the a- axis with an angle $\psi _{a}$ and generator $J_{a}$. $\left(
2\right) $ As the unit $\vec{n}_{0}$ is an eigen vector of $e^{-i\gamma
J_{z}}$; \ it follows that $\vec{n}$ reduces to $e^{-i\alpha
J_{z}}e^{-i\beta J_{y}}\vec{n}_{0}$; this generic vector obeys as well the
constraint $\left \vert \vec{n}\right \vert \equiv \left \vert \mathbf{n}%
\right \vert =1$ and is solved as follows%
\begin{equation}
\mathbf{n}=\left( \sin \beta \cos \alpha ,\sin \beta \sin \alpha ,\cos \beta
\right)
\end{equation}%
with $0\leq \alpha \leq 2\pi $ and $0\leq \beta \leq \pi $; they
parameterise the unit 2-sphere $\mathbb{S}_{\left( \mathbf{n}\right) }^{2}$
which is isomorphic to $SU\left( 2\right) /U\left( 1\right) $; the missing
angle $\gamma $ parameterises a circle $\mathbb{S}_{\left( \mathbf{n}\right)
}^{1},$ isomorphic to $U\left( 1\right) ,$ that is fibred over $\mathbb{S}%
_{\left( \mathbf{n}\right) }^{2}$.$\  \left( 3\right) $ the coherent spin
state representation gives a bridge between quantum spin operator and its
classical description; it relies on thinking of the average $<\mathbf{\hat{S}%
>}$ in terms of the classical vector $\vec{S}_{0}=\hbar S\vec{n}_{0}$
considered above ( $\vec{S}_{0}$ $\leftrightarrow $\ HWS $\left \vert
S_{0},S_{0}\right \rangle $). In this regards, recall that the $\hat{S}_{a}$
acts on classical 3-vectors $V_{b}$ through its 3$\times $3 matrix
representation like $[\hat{S}_{a},V_{b}]=-\hbar \left( J_{c}\right)
_{ab}V^{c}$ with $\left( J_{c}\right) _{ab}$ given by $-i\varepsilon _{abc}$%
; these $J_{c}$'s are precisely the generators of the $SU\left( 2\right) $
matrix representation $R\left( \alpha ,\beta ,\gamma \right) $; by replacing
$V_{b}$ by the operator $\hat{S}_{b}$, one discovers the SU$\left( 2\right) $
spin algebra $[\hat{S}_{a},\hat{S}_{b}]=i\hbar \varepsilon _{abc}\hat{S}^{c}$%
. Notice also that the classical spin vector $\vec{S}=\hbar S\vec{n}$ can be
also put in correspondence with the usual magnetic moment $\vec{\mu}=-%
\mathrm{\gamma }\vec{S}$ (with $\mathrm{\gamma }=\frac{ge}{2m}$\ the
gyromagnetic ratio); thus leading to $\vec{\mu}=\left \vert \mathbf{\mu }%
\right \vert \vec{n}$. So, the magnetization vector describes (up to a sign)
a coherent spin state with amplitude $\hbar S\mathrm{\gamma }$; and a
(opposite) time dependent direction $\vec{n}\left( t\right) $ parameterizing
the 2-sphere $\mathbb{S}_{\left( \mathbf{n}\right) }^{2}$.
\begin{equation}
n_{x}^{2}\left( t\right) +n_{y}^{2}\left( t\right) +n_{z}^{2}\left( t\right)
=1\qquad \Leftrightarrow \qquad \left \vert \vec{n}\left( t\right) \right
\vert =1  \label{0}
\end{equation}%
For explicit calculations, this unit 2-sphere equation will be often
expressed like $n^{a}n_{a}=1$; this relation leads in turns to the property $%
n^{a}dn_{a}=0$ (indicating that $\vec{n}$ and $d\vec{n}$ are normal
vectors); by implementing time, the variation $\mathbf{n}.d\mathbf{n}$ gets
mapped into $\mathbf{n}.\mathbf{\dot{n}}=0$ teaching us that the velocity $%
\mathbf{\dot{n}}$ is carried by $\mathbf{u}$ and $\mathbf{v}$; two normal
directions to $\mathbf{n}$ with components;%
\begin{equation}
u_{a}=\left( \cos \beta \cos \alpha ,\cos \beta \sin \alpha ,-\sin \beta
\right) \quad ,\quad v_{a}=\left( -\sin \alpha ,\cos \alpha ,0\right)
\label{1}
\end{equation}%
and from which we learn that $dn_{a}=u_{a}d\beta +v_{a}\sin \beta d\alpha $,
[($\mathbf{u,v,n}$) form an orthogonal vector triad). So, the dynamics of $%
\mu _{a}$ (and that of $-\vec{S}$) is brought to the dynamics of the unit $%
n_{a}$ governed by a classical Hamiltonian H$\left[ n_{a}\left( \alpha
,\beta \right) \right] $. The resulting time evolution is given by the so
called Landau-Lifshitz (LL) equation \textrm{\cite{6a}}; it reads as $\frac{%
dn_{a}}{dt}=-\frac{\mathrm{\gamma }}{\left \vert \mathbf{\mu }\right \vert }%
\varepsilon _{abc}\left( \partial ^{b}H\right) n^{c}$ with $\partial ^{b}H=%
\frac{\partial H}{\partial n_{b}}.$ By using the relations $d\beta
=u^{a}dn_{a}$ and $\sin \beta d\alpha =v^{a}dn_{a}$ with $u_{a}=\frac{%
\partial n_{a}}{\partial \beta }$; and $v_{a}\sin \beta =\frac{\partial n_{a}%
}{\partial \alpha }$; as well as the expressions $\varepsilon
_{abc}u^{a}n^{c}=v_{b}$ and $\varepsilon _{abc}v^{a}n^{c}=-u_{b}$, the above
LL equation splits into two time evolution equations $\frac{d\beta }{dt}=-%
\mathrm{\gamma }v_{b}\left( \partial ^{b}H\right) $ and $\sin \beta \frac{%
d\alpha }{dt}=\mathrm{\gamma }u_{b}\left( \partial ^{b}H\right) $. These
time evolutions can be also put into the form%
\begin{equation}
\sin \beta \frac{d\beta }{dt}+\mathrm{\gamma }\frac{\partial H}{\partial
\alpha }=0\quad ,\quad \sin \beta \frac{d\alpha }{dt}-\mathrm{\gamma }\frac{%
\partial H}{\partial \beta }=0  \label{2}
\end{equation}%
and can be identified with the Euler- Lagrange equations following from the
variation $\delta \mathcal{S}=0$ of an action $\mathcal{S}=\int Ldt$. Here,
the Lagrangian is related to the Hamiltonian like $L=L_{B}-H\left[
n_{a}\left( \alpha ,\beta \right) \right] $ where $L_{B}$ is the Berry term
\textrm{\cite{7a}} known to have the form \TEXTsymbol{<}$\mathbf{n}$%
\TEXTsymbol{\vert}$\mathbf{\dot{n}}$\TEXTsymbol{>}; this relation can be
compared with the well known Legendre transform $p\dot{q}-H\left( q,p\right)
$. For later interpretation, we scale this hamiltonian as $\hbar S\mathrm{%
\gamma }H$ such that the spin lagrangian takes the form $L_{spin}=L_{B}-%
\hbar S\mathrm{\gamma }H$. To determine $L_{B}$, we identify the equations (%
\ref{2}) with the extremal variation $\delta \mathcal{S}/\delta \beta =0$
and $\delta \mathcal{S}/\delta \alpha =0$. Straightforward calculations
leads to%
\begin{equation}
L_{B}=-\hbar S\left( 1-\cos \beta \right) \frac{d\alpha }{dt}  \label{bl}
\end{equation}%
showing that $\alpha $ and $\beta $ form a conjugate pair. By substituting $%
\sin \beta \frac{d\alpha }{dt}=v^{a}\frac{dn_{a}}{dt}$ back into above $%
L_{B} $, we find that the Berry term has the form of Aharonov-Bohm coupling $%
L_{AB}=q_{e}A^{a}\frac{dn_{a}}{dt}$ with magnetic potential vector $A^{a}$
given by $A_{a}=\frac{\hbar S}{q_{e}}\frac{\left( 1-\cos \beta \right) }{%
\sin \beta }v_{a}.$ However, this potential vector is suggestive as it has
the same form as the potential vector $\boldsymbol{A}^{{\small (monopole)}}=%
\frac{\hbar S}{q_{e}r}\frac{\left( 1-\cos \beta \right) }{\sin \beta }%
\mathbf{v}$ of a magnetic monopole. The curl of this potential is given by $%
\boldsymbol{\vec{B}}=q_{m}\frac{\vec{r}}{r^{3}}$ with magnetic charge $%
q_{m}=-\frac{\hbar S}{q_{e}}$ located at the centre of the 2-sphere; the
flux $\Phi $ of this field through the unit sphere is then equal to $-4\pi
\frac{\hbar S}{q_{e}}$; and reads as $-2S\Phi _{0}$ with a unit flux quanta $%
\Phi _{0}=\frac{h}{q_{e}}$ as indicated by the value $S=1/2$. So, because $%
2S=-n$ is an integer, it results that the flux is quantized as $\Phi =n\Phi
_{0}.$

\section{Magnetic solitons in lower dimensions}

In previous section, we have considered the time dynamics of coherent spin
states with amplitude $\hbar S$ and direction described by $\vec{n}\left(
t\right) $ as depicted by the Figure \textbf{\ref{spin1}}-(a); this is a
3-vector having with no space coordinate dependence, $\func{grad}\vec{n}=0$;
and as such it can be interpreted as a $\left( 1+0\right) D$ vector field;
that is a vector belonging to $\mathbb{R}^{1,d}$ with $d=0$ (no space
direction). In this section, we first turn on 1d space coordinate $x$ and
promotes the old unit- direction $\vec{n}\left( t\right) $ to a $\left(
1+1\right) $D field $\vec{n}\left( t,x\right) $. After that, we turn on two
space directions $\left( x,y\right) $; thus leading to $\left( 1+2\right) $D
field $\vec{n}\left( t,x,y\right) ;$ a picture is depicted by the Figure
\textbf{\ref{spin1}}-(b). To deal with the dynamics of these local fields
and their topological properties, we use the field theory method while
focussing on particular solitons; namely the 1d kinks and the 2d skyrmions.
In this extension, one encounters two types of spaces: $\left( 1\right) $
the target space $\mathbb{R}_{\mathbf{n}}^{3}$ parameterised by $%
n_{a}=\left( n_{1},n_{2},n_{3}\right) $ with Euclidian metric $\delta _{ab}$
and topological Levi-Civita $\varepsilon _{abc}$. $\left( 2\right) $ the
spacetime $\mathbb{R}_{\xi }^{1,1}$ parameterised by $\xi ^{\mu }=\left(
t,x\right) ,$ concerning the 1d kink evolution; and the spacetime $\mathbb{R}%
_{\xi }^{1,2}$ parameterised by $\xi ^{\mu }=\left( t,x,y\right) $,
regarding the 2d skyrmions dynamics. As we have two kinds of evolutions;
time and space; we denote the time variable by $\xi ^{0}=t$; and the space
coordinates by $\xi ^{i}=\left( x,y\right) $. Moreover, the homologue of the
tensors $\delta _{ab}$ and $\varepsilon _{abc}$ are respectively given by
the usual Lorentzian spacetime metric $g_{\mu \nu }$, with signature like $%
g_{\mu \nu }\xi ^{\mu }\xi ^{\nu }=x^{2}+y^{2}-t^{2}$, and the spacetime
Levi-Civita $\epsilon _{\mu \nu \rho }$ with $\epsilon _{012}=1$.

\subsection{One space dimensional solitons}

In $\left( 1+1\right) $D spacetime, the local coordinates parameterising $%
\mathbb{R}_{\left( \xi \right) }^{1,1}$ are given by $\xi ^{\mu }=\left(
t,x\right) $; so the metric is restricted to $g_{\mu \nu }\xi ^{\mu }\xi
^{\nu }=x^{2}-t^{2}$. The field variable $n^{a}\left( \xi \right) $ has in
general three components $\left( n_{1},n_{2},n_{3}\right) $ as described
previously; but in what follows, we will simplify a little bit the picture
by setting $n_{3}=0$; thus leading to a magnetic 1d soliton with two
component field variable $\mathbf{n}=\left( n_{1},n_{2}\right) $ satisfying
the constraint equation $\mathbf{n.n}=1$ at each point of spacetime. As this
constraint relation plays an important role in the construction, it is
interesting to express it as $n_{a}n^{a}=1$. Before describing the
topological properties of one space dimensional solitons (kinks), we think
it interesting to begin by giving first some useful features; in particular
the \textrm{three} following ones. $\left( 1\right) $ The constraint $\left(
n_{1}\right) ^{2}+\left( n_{2}\right) ^{2}=1$ is invariant $SO\left(
2\right) _{\mathbf{n}}$ rotations acting as $n^{\prime a}=\mathcal{R}%
_{b}^{a}n^{b}$ with orthogonal rotation matrix
\begin{equation}
\mathcal{R}_{b}^{a}=\left(
\begin{array}{cc}
\cos \psi & \sin \psi \\
-\sin \psi & \cos \psi%
\end{array}%
\right) \qquad ,\qquad \mathcal{R}^{T}\mathcal{R}=I
\end{equation}%
The constraint $n_{a}n^{a}=1$ can be also presented like $\bar{N}N=1$ with $%
N $ standing for the complex field $n_{1}+in_{2}$ that reads also like $%
e^{i\alpha }$. In this complex notation, the symmetry of the constraint is
given by the phase change acting as $N\rightarrow UN$ with $U=e^{i\psi }$
and corresponding to the shift $\alpha \rightarrow \alpha +\psi $. Moreover
the correspondence $\left( n_{1},n_{2}\right) \leftrightarrow n_{1}+in_{2}$
describes precisely the well known isomorphisms $SO\left( 2\right) \sim
U\left( 1\right) \sim \mathbb{S}_{\left( \mathbf{n}\right) }^{1}$ where $%
\mathbb{S}_{\left( \mathbf{n}\right) }^{1}$ is a circle; it is precisely the
equatorial circle of the 2-sphere $\mathbb{S}_{\left( \mathbf{n}\right)
}^{2} $ considered in previous section. $\left( 2\right) $ As for Eq(\ref{1}%
), the constraint $n_{a}n^{a}=1$ leads to $n_{a}dn^{a}=0$; and so describes
a rotational movement encoded in the relation $dn^{a}=\varepsilon ^{ab}n_{b}$
where $\varepsilon ^{ab}$ is the standard 2D antisymmetric tensor with $%
\varepsilon ^{21}=\varepsilon _{12}=1$; this $\varepsilon _{ab}$ is related
to the previous 3D Levi-Civita like $\varepsilon _{zab}$. Notice also that
the constraint $n_{a}n^{a}=1$ implies moreover that $dn_{2}=-\frac{n_{n}}{%
n_{2}}dn_{1};$ and consequently the area $dn_{1}\wedge dn_{2}$, to be
encountered later on, vanishes identically. In this regards, recall that we
have the following transformation%
\begin{equation}
dn_{1}\wedge dn_{2}=\mathfrak{J}dt\wedge dx\quad ,\quad \mathfrak{J}%
=\epsilon ^{\mu \nu }\partial _{\mu }n_{1}\partial _{\nu }n_{2}  \label{ct}
\end{equation}%
where $\epsilon ^{\mu \nu }$ is the antisymmetric tensor in 1+1 spacetime,
and $\mathfrak{J}$ is the Jacobian of the transformation $\left( t,x\right)
\rightarrow \left( n_{1},n_{2}\right) $. $\left( 3\right) $ The condition $%
n_{a}n^{a}=1$ can be dealt in two manners; either by inserting it by help of
a Lagrange multiplier; or by solving it in term of a free angular variable
like $n_{a}=\left( \cos \alpha ,\sin \alpha \right) $ from which we deduce
the normal direction $u^{a}=\frac{dn^{a}}{d\alpha }$ reading as $%
u_{a}=\left( -\sin \alpha ,\cos \alpha \right) $. In term of the complex
field; we have $N=e^{i\alpha }$ and $\bar{N}dN=id\alpha $. Though
interesting, the second way of doing hides an important property in which we
are interested in here namely the non linear dynamics and the topological
symmetry.

\subsubsection{Constrained dynamics}

The classical spacetime dynamics of $n^{a}\left( \xi \right) $ is described
by a field action $\mathcal{S}=\int dtL$ with Lagrangian $L=\int dx\mathcal{L%
}$ and density $\mathcal{L}$; this field density is given by $-\frac{1}{2}%
\left( \partial _{\mu }n_{a}\right) \left( \partial ^{\mu }n^{a}\right)
-V\left( n\right) -\Lambda \left( n^{a}n_{a}-1\right) $ with $\partial _{\mu
}=\frac{\partial }{\partial \xi ^{\mu }}$; it reads in terms of the
Hamiltonian density as follows%
\begin{equation}
\mathcal{L}=\pi ^{a}\dot{n}_{a}-\mathcal{H}  \label{b3}
\end{equation}%
where $\pi ^{a}=\frac{\partial \mathcal{L}}{\partial \dot{n}_{a}}$. In the
above Lagrangian density, the auxiliary field $\Lambda \left( \xi \right) $
(no Kinetic term) is a Lagrange multiplier carrying the constraint relation $%
n_{a}n^{a}=1$. The $V\left( n\right) $ is a potential energy density which
play an important role for describing 1d kinks with finite size. Notice also
that the variation $\frac{\delta \mathcal{S}}{\delta \Lambda }=0$ gives
precisely the constraint $n_{a}n^{a}=1$ while the $\frac{\delta \mathcal{S}}{%
\delta n^{a}}=0$ gives the spacetime dynamics of $n^{a}$ described by the
spacetime equation $\partial _{\mu }\partial ^{\mu }n^{a}-\frac{\partial V}{%
\partial n^{a}}-\Lambda n^{a}=0$. By substituting $n_{a}=\left( \cos \alpha
,\sin \alpha \right) $, we obtain $\mathcal{L}=-\frac{1}{2}\left( \partial
_{\mu }\alpha \right) \left( \partial ^{\mu }\alpha \right) -V\left( \alpha
\right) $. If setting $V\left( \alpha \right) =0$, we end up with the free
field equation $\partial _{\mu }\partial ^{\mu }\alpha =0$ that expands like
$\left( \partial _{x}^{2}-\partial _{t}^{2}\right) \alpha =0$; it is
invariant under spacetime translations with conserved current symmetry $%
\partial ^{\mu }T_{\mu \nu }=0$ with $T_{\mu \nu }$\ standing for the energy
momentum tensor given by the 2$\times $2 symmetric matrix $\partial _{\mu
}\alpha \partial _{\nu }\alpha +g_{\mu \nu }\mathcal{L}$. The energy density
$T_{00}$ is given by $\frac{1}{2}\left( \partial _{t}\alpha \right) ^{2}+%
\frac{1}{2}\left( \partial _{x}\alpha \right) ^{2}$ and the momentum density
$T_{10}$ reads as $\partial _{x}\varphi \partial _{x}\varphi $. Focussing on
$T_{00}$, the conserved energy $E$ reads then as follows%
\begin{equation}
E=\frac{1}{2}\int_{-\infty }^{+\infty }dx\left[ \left( \partial _{t}\alpha
\right) ^{2}+\left( \partial _{x}\alpha \right) ^{2}\right] \geq 0  \label{e}
\end{equation}%
with minimum corresponding to constant field ($\alpha =cte$). Notice that
general solutions of $\partial _{\mu }\partial ^{\mu }\alpha =0$ are given
by arbitrary functions $f\left( x\pm t\right) $; they include oscillating
and non oscillating functions. A typical non vibrating solution that is
interesting for the present study is the solitonic solution given (up to a
constant $c$) by the following expression%
\begin{equation}
\varphi \left( t,x\right) =\pi \tanh \left( \frac{x+t}{\lambda }\right)
\label{fi}
\end{equation}%
where $\lambda $\ is a positive parameter representing the width where the
soliton $\alpha \left( t,x\right) $ acquires a significant variation. Notice
that for a given $t$, the field varies from $\alpha \left( t,-\infty \right)
=-\pi $ to $\alpha \left( t,+\infty \right) =\pi $ regardless the value of $%
\lambda $. These limits are related to each other by a period $2\pi $.

\subsubsection{Topological current and charge}

To start, notice that as far as conserved symmetries of (\ref{b3}) are
concerned, there exists an exotic invariance generated by a conserved $%
J_{\mu }\left( t,x\right) $ going beyond the spacetime translations
generated by the energy momentum tensor $T_{\mu \nu }$. The conserved
spacetime current $J_{\mu }=\left( J_{0},J_{1}\right) $ of this exotic
symmetry can be introduced in two different, but equivalent, manners; either
by using the free degree of freedom $\alpha $; or by working with the
constrained field $n^{a}.$ In the first way, we think of the charge density $%
J_{0}$ like $\frac{1}{2\pi }\partial _{1}\alpha $ and of the current density
as $J_{1}=-\frac{1}{2\pi }\partial _{0}\alpha $. This conserved current is a
topological $\left( 1+1\right) D$ spacetime vector $J_{\mu }$ that is
manifestly conserved; this feature follows from the relation between $J_{\mu
}$ and the antisymmetric $\epsilon _{\mu \nu }$ as follows \textrm{\cite{3a1}%
,}
\begin{equation}
J_{\mu }=\frac{1}{2\pi }\epsilon _{\mu \nu }\partial ^{\nu }\alpha
\label{jm}
\end{equation}%
Because of the $\epsilon _{\mu \nu }$; the continuity relation $\partial
^{\mu }J_{\mu }=\frac{1}{2\pi }\epsilon _{\mu \nu }\partial ^{\mu }\partial
^{\nu }\alpha $ vanishes identically due to the antisymmetry property of $%
\epsilon _{\mu \nu }$. The particularity of the above conserved $J_{\mu }$
is its topological nature; it is due to the constraint $n_{a}n^{a}=1$
without recourse to the solution $n_{a}=\left( \cos \alpha ,\sin \alpha
\right) $. Indeed, Eq(\ref{jm}) can be derived by computing the Jacobian $%
\mathfrak{J}=\det (\frac{\partial n^{a}}{\partial \xi ^{\mu }})$ of the
mapping from the 2d spacetime coordinates $\left( t,x\right) $ to the target
space fields ($n_{1},n_{2}$). Recall that the spacetime area $dt\wedge dx$
can be written in terms of $\epsilon _{\mu \nu }$ like $\frac{1}{2}\epsilon
_{\mu \nu }d\xi ^{\mu }\wedge d\xi ^{\nu }$ and, similarly, the target space
area $dn_{1}\wedge dn_{2}$ can be expressed in terms of as $\varepsilon
_{ab} $ follows $\frac{1}{2}\varepsilon _{ab}dn^{a}\wedge dn^{b}$. The
Jacobian $\mathfrak{J}$ is precisely given by (\ref{ct}); and can be
presented into a covariant form like $\mathfrak{J}=\frac{1}{2}\varepsilon
^{\mu \nu }\partial _{\mu }n^{a}\partial _{\nu }n^{b}\varepsilon _{ab}$.
This expression of the Jacobian $\mathfrak{J}$ captures important
informations; in particular the \textrm{three following ones. }$\left(
1\right) $ It can be expressed as a total divergence like $\partial _{\mu
}\left( \pi J^{\mu }\right) $ with spacetime vector
\begin{equation}
J^{\mu }=\frac{1}{2\pi }\varepsilon ^{\mu \nu }n^{a}\partial _{\nu
}n^{b}\varepsilon _{ab}
\end{equation}%
and where $\frac{1}{\pi }$ is a normalisation; it is introduced for the
interpretation of the topological charged as just the usual winding number
of the circle [encoded in the homotopy group relation $\pi _{1}\left(
\mathbb{S}^{1}\right) =\mathbb{Z}$]. $\left( 2\right) $ Because of the
constraint $dn_{2}=-\frac{n_{n}}{n_{2}}dn_{1}$ following from $n_{a}n^{a}=1$%
, the Jacobian $\mathfrak{J}$ vanishes identically; thus leading to the
conservation law $\partial _{\mu }J^{\mu }=0$; i.e $\mathfrak{J}=0$ and then
$\partial _{\mu }J^{\mu }=0$. $\left( 3\right) $ The conserved charge $Q$
associated with the topological current is given by $\int_{-\infty
}^{+\infty }dxJ^{0}\left( t,x\right) $; it is time independent despite the
apparent t- variable in the integral ($dQ/dt=0$). By using (\ref{jm}), this
charge reads also as $\frac{1}{2\pi }\int_{-\infty }^{+\infty }dx\partial
_{x}\alpha \left( t,x\right) $ and after integration leads to
\begin{equation}
Q=\frac{1}{2\pi }\left[ \alpha \left( t,\infty \right) -\alpha \left(
t,-\infty \right) \right]
\end{equation}%
Moreover, seen that $\alpha \left( t,\infty \right) $ is an angular variable
parameterising $\mathbb{S}_{\mathbf{n}}^{1}$; it may be subject to a
boundary condition like for instance the periodic $\alpha \left( t,\infty
\right) =\alpha \left( t,-\infty \right) +2\pi N$ with N an integer; this
leads to an integral topological charge $Q=N$ interpreted as the winding
number of the circle. In this regards, notice that: $\left( i\right) $ the
winding interpretation can be justified by observing that under
compactification of the space variable $x$, the infinite space line $\mathbb{%
R}_{x}=\left] -\infty ,+\infty \right[ $ gets mapped into a circle $\mathbb{S%
}_{\left( x\right) }^{1}$ with angular coordinate $-\pi \leq \varphi \leq
\pi $; so, the integral $\frac{1}{2\pi }\int_{-\infty }^{+\infty }dx\partial
_{x}\alpha \left( t,x\right) $ gets replaced by $\frac{1}{2\pi }\int_{-\pi
}^{+\pi }d\varphi \frac{\partial \alpha }{\partial \varphi }$; and then the
mapping $\alpha _{t}:\varphi \rightarrow \alpha \left( t,\varphi \right) $
is a mapping between two circles namely $\mathbb{S}_{\left( x\right)
}^{1}\rightarrow \mathbb{S}_{\left( \mathbf{n}\right) }^{1}$; the field $%
\alpha \left( t,\varphi \right) $ then describes a soliton (one space
extended object) wrapping the circle $\mathbb{S}_{\left( x\right) }^{1}$ N
times; this propery is captured by $\pi _{\mathbb{S}_{\left( x\right) }^{1}}(%
\mathbb{S}_{\left( \mathbf{n}\right) }^{1})=\mathbb{Z}$, a homotopy group
property \textrm{\cite{1b}}. $\left( ii\right) $ The charge $Q$ is
independent of the Lagrangian of the system as it follows completely from
the field constraint without any reference to the field action. $\left(
iii\right) $ Under a scale transformation $\xi ^{\prime }=\xi /\lambda $
with a scaling parameter $\lambda >0$, the topological charge of the field (%
\ref{fi}) is invariant; but its total energy (\ref{e}) get scaled as follows%
\begin{equation}
Q^{\prime }=Q\qquad ,\qquad E^{\prime }=\frac{1}{\lambda }E
\end{equation}%
This energy transformation shows that stable solitons with minimal energy
correspond to $\lambda \rightarrow \infty $; and then to a trivial soliton
spreading along the real axis. However, one can have non trivial solitonic
configurations that are topologically protected and energetically stable
with non diverging $\lambda $. This can done by turning on an appropriate
potential energy density $V\left( \mathbf{n}\right) $ in Eq(\ref{b3}). An
example of such potential is the one given by $\frac{g}{8}\left(
n_{1}^{4}+n_{2}^{4}-1\right) $, with positive $g=M^{2}$, breaking $SO\left(
2\right) _{\mathbf{n}}$; by using the constraint $n_{1}^{2}+n_{2}^{2}=1$, it
can be put $\frac{g}{4}n_{1}^{2}n_{2}^{2}$. In terms of the angular field $%
\alpha $, it reads as $V\left( \alpha \right) =\frac{g}{16}\left( 1-\cos
4\alpha \right) $ leading to the well known sine-Gordon equation \textrm{%
\cite{2b,2b1}} namely $\partial _{\mu }\partial ^{\mu }\alpha -\frac{g}{4}%
\sin 4\alpha =0$ with the symmetry property $\alpha \rightarrow \alpha +%
\frac{\pi }{2}$. So, the solitonic solution is periodic with period $\frac{%
\pi }{2}$; that is the quarter of the old 2$\pi $ period of the free field
case. For static field $\alpha \left( x\right) $, the sine Gordon equation
reduces to $\frac{d^{2}\alpha }{dx^{2}}-\frac{M^{2}}{4}\sin 4\alpha =0$; its
solution for $M>0$ is given by $\arctan \left[ \exp Mx\right] $ representing
a sine- Gordon field evolving from $0$ to $\frac{\pi }{2}$ and describing a
kink with topological charge $Q=\frac{1}{4}.$ For $M<0$, the soliton is an
anti-kink evolving from $\frac{\pi }{2}$ to $0$ with charge $Q=-\frac{1}{4}.$
Time dependent solutions can be obtained by help of boost transformations $%
x\rightarrow \frac{x\pm vt}{\sqrt{1-v^{2}}}$.

\subsection{Skyrmions in 2d space dimensions}

In this subsection, we investigate the topological properties of 2d
Skyrmions by extending the field theory study we have done above for 1d
kinks to two space dimensions. For that, we proceed as follows: First, we
turn on the component $n_{3}$ so that the skyrmion field $\mathbf{n}$ is a
real 3-vector with three components $\left( n_{1},n_{2},n_{3}\right) $
constrained as in Eqs(\ref{0}-\ref{1}).
\begin{figure}[tbph]
\begin{center}
\includegraphics[width=10cm]{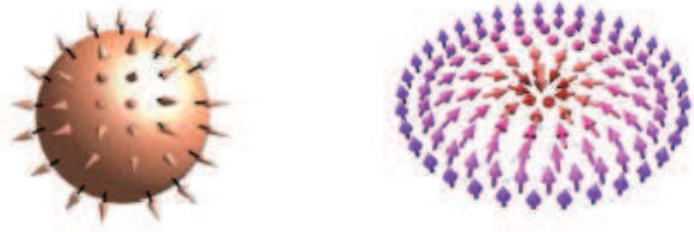}
\end{center}
\par
\vspace{-1cm}
\caption{{\protect \small On left: a spin configuration with }$%
n_{1}^{2}+n_{2}^{2}+n_{3}^{2}=1${\protect \small \ dispatched on a 2-sphere.
On right: a two space dimensional magnetic skyrmion given by the
stereographic projection of }$\mathbb{S}^{{\protect \small 2}}$%
{\protect \small \ to plane.}}
\label{hedgehog}
\end{figure}
Second, here we have $\mathbf{n}=\mathbf{n}\left( t,x,y\right) $; that is a
3-component field living in the $\left( 2+1\right) $ space time with
Lorentzian metric and coordinates $\xi ^{\mu }=\left( t,x,y\right) $. This
means that $d\mathbf{n}=(\partial _{\mu }\mathbf{n})d\xi ^{\mu }$;
explicitly $d\mathbf{n}=\frac{\partial \mathbf{n}}{\partial t}dt+\frac{%
\partial \mathbf{n}}{\partial x}dx+\frac{\partial \mathbf{n}}{\partial y}dy.$

\subsubsection{Dzyaloshinskii-Moriya potential}

The field action $\mathcal{S}_{3D}=\int dtL_{3D}$ describing the space time
dynamics of $\mathbf{n}\left( t,x,y\right) $ has the same structure as eq(%
\ref{b3}); except that here the Lagrangian $L_{3D}$ involves two space
variable like $\int dxdy\mathcal{L}_{3D}$ and the density $\mathcal{L}_{3D}=-%
\frac{1}{2}\left( \partial _{\mu }\mathbf{n}\right) ^{2}-V\left( \mathbf{n}%
\right) -\Lambda \left( \mathbf{n.n}-1\right) $; this is a function of the
constrained 3-vector $\mathbf{n}$ and its space time gradient $\partial
_{\mu }\mathbf{n}$; it reads in term of the Hamiltonian density as follows.
\begin{equation}
\mathcal{L}_{3D}=\mathbf{\pi }\mathbf{.\dot{n}}-\mathcal{H}_{3D}\left(
\mathbf{n}\right)  \label{3d}
\end{equation}%
In this expression, the $\mathcal{H}_{3D}\left( \mathbf{n}\right) $ is the
continuous limit of a lattice Hamiltonian H$_{latt}([n^{a}\left( \mathbf{r}%
_{\mu }\right) ]$ involving, amongst others, the Heisenberg term, the
Dyaloshinskii- Moriya (DM) interaction and the Zeeman coupling. The $V\left(
\mathbf{n}\right) $ in the first expression of $\mathcal{L}_{3D}$ is the
scalar potential energy density; it models the continuous limit of the
interactions that include the DM and Zeeman ones [see eq(1.22) for its
explicit relation]. The field $\Lambda \left( \xi \right) $ is an auxiliary
3D spacetime field; it is a Lagrange multiplier that carries the constraint $%
\mathbf{n.n}=1$ which plays the same role as in \textrm{subsection 3.1}. By
variating this action with respect to the fields $\mathbf{n}$ and $\Lambda $%
; we get from $\frac{\delta \mathcal{S}_{3D}}{\delta \Lambda }=0$ precisely
the field constraint $\mathbf{n.n}=1$; and from $\frac{\delta \mathcal{S}%
_{3D}}{\delta \mathbf{n}}=0$ the following Euler-Lagrange equation $\square
\mathbf{n}=\frac{\partial V}{\partial \mathbf{n}}+\Lambda \mathbf{n}.$ For
later use, we express this field equation like%
\begin{equation}
\partial ^{\mu }\partial _{\mu }n_{a}=\frac{\partial V}{\partial n^{a}}%
+\Lambda n_{a}  \label{nv}
\end{equation}%
The interest into this (\ref{nv}) is twice; first it can be put into the
equivalent form $\partial ^{\mu }\partial _{\mu }n_{a}=\varepsilon _{abc}%
\mathcal{D}^{b}n^{c}$ where $\mathcal{D}^{b}$ is an operator acting on $%
n^{c} $ to be derived later on [see eq(\ref{da})]; and second, it can be
used to give the relation between the scalar potential and the operator $%
\mathcal{D}^{b}$. To that purpose, we start by noticing that there are two
manners to deal with the field constraint $n^{a}n_{a}=1$; either by using
the Lagrange multiplier $\Lambda $; or by solving it in terms of two angular
field variables as given by eq(\ref{1}). In the second case, we have the
triad $n_{a}=\left( \sin \beta \cos \alpha ,\sin \beta \sin \alpha ,\cos
\beta \right) $ and%
\begin{equation}
u_{a}=\left( \cos \beta \cos \alpha ,\cos \beta \sin \alpha ,-\sin \beta
\right) \quad ,\quad v_{a}=\left( -\sin \alpha ,\cos \alpha ,0\right)
\label{nuv}
\end{equation}%
but now $\beta =\beta \left( t,x,y\right) $ and $\alpha =\alpha \left(
t,x,y\right) $ with $0\leq \beta \leq \pi $ and $0\leq \alpha \leq 2\pi .$
Notice also that the variation of the filed constraint leads to $%
n_{a}dn^{a}=0$ teaching us interesting informations, in particular the two
following useful ones. $\left( 1\right) $ the movement of $n_{a}$ in the
target space is a rotational movement; and so can be expressed like
\begin{equation}
dn_{a}=\varepsilon _{abc}\omega ^{b}n^{c}\quad \Leftrightarrow \quad d%
\mathbf{n}=\mathbf{\omega }\wedge \mathbf{n\quad \Leftrightarrow \quad
\omega }\sim \mathbf{n}\wedge d\mathbf{n}  \label{me}
\end{equation}%
where the 1-form $\omega ^{b}$ is the rotation vector to be derived below.
By substituting (\ref{me}) back into $n^{a}dn_{a},$\ we obtain $\varepsilon
_{bca}\omega ^{b}n^{c}n^{a}$ which vanishes identically due to the property $%
\varepsilon _{bca}n^{c}n^{a}=0$. $\left( 2\right) $ Having two degrees of
freedom $\alpha $ and $\beta $, we can expand the differential $dn_{a}$ like
$u_{a}d\beta +v_{a}\sin \alpha d\alpha $ with the two vector fields $u_{a}=%
\frac{\partial n_{a}}{\partial \beta }$ and $v_{a}=\frac{\partial n_{a}}{%
\partial \alpha }$ as given above. Notice that the three unit fields $\left(
\mathbf{n},\mathbf{u},\mathbf{v}\right) $ plays an important role in this
study; they form a vector basis of the field space; they obey the usual
cross products namely $\mathbf{n}=\mathbf{u}\wedge \mathbf{v}$ and its
homologue which given by cyclic permutations; for example,
\begin{equation}
u_{a}=\varepsilon _{abc}v^{b}n^{c}\quad ,\quad v_{a}=-\varepsilon
_{abc}u^{b}n^{c}  \label{em}
\end{equation}%
Putting these eqs(\ref{em}) back into the expansion of $dn_{a}$ in terms of $%
d\alpha ,$ $d\beta $; and comparing with Eq(\ref{me}), we end up with the
explicit expression of the 1-form angular "speed" vector $\omega ^{b}$; it
reads as follows $\omega ^{b}=v^{b}d\beta -u^{b}\sin \alpha d\alpha $.
Notice that by using the space time coordinates $\xi $, we can also express
eq(\ref{me}) like $\partial _{\mu }n_{a}=\varepsilon _{abc}\omega _{\mu
}^{b}n^{c}$ with $\omega _{\mu }^{b}$ given by $v^{b}(\partial _{\mu }\beta
)-u^{b}\sin \alpha (\partial _{\mu }\alpha ).$ From this expression, we can
compute the Laplacian $\partial ^{\mu }\partial _{\mu }n_{a}$; which, by
using the above relations; is equal to $\varepsilon _{abc}\partial ^{\mu
}\left( \omega _{\mu }^{b}n^{c}\right) $ reading explicitly as $\varepsilon
_{abc}[(\partial ^{\mu }\omega _{\mu }^{b})n^{c}+\omega _{\mu }^{b}(\partial
^{\mu }n^{c})]$ or equivalently like $\partial ^{\mu }\partial _{\mu
}n_{a}=\varepsilon _{abc}\mathcal{D}^{b}n^{c}$ with operator $\mathcal{D}%
^{b}=\omega _{\mu }^{b}\partial ^{\mu }+(\partial ^{\mu }\omega _{\mu }^{b})$%
. Notice that the above operator has an interesting geometric
interpretation; by factorising $\omega _{\mu }^{b}$, we can put it in the
form $\omega _{\mu }^{d}\left( \mathcal{D}^{\mu }\right) _{d}^{b}$ where $%
\left( \mathcal{D}^{\mu }\right) _{d}^{b}$ appears as a gauge covariant
derivative $\left( \mathcal{D}^{\mu }\right) _{d}^{b}=\delta
_{d}^{b}\partial ^{\mu }+\left( A^{\mu }\right) _{d}^{b}$ with a non trivial
gauge potential $\left( A^{\mu }\right) _{d}^{b}$ given by $\omega _{d}^{\mu
}(\partial ^{\nu }\omega _{\nu }^{b}).$ Comparing with (\ref{nv}) with $%
\partial ^{\mu }\partial _{\mu }n_{a}=\varepsilon _{abc}\mathcal{D}^{b}n^{c}$%
, we obtain $\frac{\partial V}{\partial n^{a}}=\varepsilon _{abc}\mathcal{D}%
^{b}n^{c}-\Lambda n_{a};$ and then a scalar potential energy $V$ given by $%
\int \varepsilon _{abc}(dn^{a}\mathcal{D}^{b}n^{c})-\Lambda \int n_{a}dn^{a}$%
. The second term in this relation vanishes identically because $%
n_{a}dn^{a}=0$; thus reducing to
\begin{equation}
V=\int \varepsilon _{abc}(dn^{a}\mathcal{D}^{b}n^{c})  \label{V}
\end{equation}%
containing $\varepsilon _{abc}(n^{a}\mathcal{D}^{b}n^{c})$ as a sub-term. In
the end of this analysis, let us compare this sub-term with the $\varepsilon
_{abc}n^{b}n_{\mu _{1}\mu _{2}}^{c}\Delta ^{a\mu _{1}\mu _{2}}$ with $\Delta
^{a\mu _{1}\mu _{2}}=d_{\mu _{3}...\mu _{D-2}}^{a}\epsilon ^{\mu _{1}...\mu
_{D}}$ giving the general structure of the DM coupling (see end of \textrm{%
subsection 2.1}). For $\left( 1+2\right) $D spacetime, the general structure
of DM interaction reads $\varepsilon _{abc}d_{0}^{a}\left( n^{b}\mathbf{%
\nabla }n^{c}\right) .\mathbf{e}_{\mu \nu }\epsilon ^{0\mu \nu }$; by
setting $\mathbf{e}^{0}=\mathbf{e}_{\mu \nu }\epsilon ^{0\mu \nu }$ and $%
\mathbf{d}^{a}=d_{0}^{a}\mathbf{e}^{0}$ as well as $D^{a}=\mathbf{d}^{a}.%
\mathbf{\nabla }$, one brings it to the form $\varepsilon _{abc}\left(
n^{a}D^{b}n^{c}\right) $ which is the same as the one following from (\ref{V}%
).

\subsubsection{From kinks to 2d Skyrmions}

Here, we study the topological properties of the 2d Skyrmion with dynamics
governed by the Lagrangian density (\ref{3d}). From the expression of the $%
\left( 1+1\right) $D topological current $(J^{\mu })_{{\small 2D}}$
discussed in subsection 2.1, which reads as $\frac{1}{2\pi }\epsilon ^{\mu
\nu }n^{a}\partial _{\nu }n^{b}\varepsilon _{ab}$, one can wonder the
structure of the $\left( 1+2\right) $D topological current $(J^{\mu })_{%
{\small 3D}}$ that is associated with the 2d Skyrmion described by the
3-vector field $n_{a}\left( \xi \right) $. It is given by
\begin{equation}
\left( J^{\mu }\right) _{{\small 3D}}=\frac{1}{8\pi }\epsilon ^{\mu \nu \rho
}n^{a}\partial _{\nu }n^{b}\partial _{\nu }n^{c}\varepsilon _{abc}
\label{cs}
\end{equation}%
where $\varepsilon _{abc}$ is as before and where $\epsilon ^{\mu \nu \rho }$
is the completely antisymmetric Levi-Civita tensor in the $\left( 1+2\right)
$D spacetime. The divergence $\partial _{\mu }(J^{\mu })_{{\small 3D}}$ of
the above spacetime vector vanishes identically; it has two remarkable
properties that we want to comment before proceeding. $\left( 1\right) $ The
$\partial _{\mu }(J^{\mu })_{{\small 3D}}$ is nothing but the determinant of
the $3\times 3$ Jacobian matrix $\frac{\partial n^{a}}{\partial \xi ^{\mu }}$
relating the three field variables $n^{a}$ to the three spacetime
coordinates $\xi ^{\mu }$; this Jacobian $\det (\frac{\partial n^{a}}{%
\partial \xi ^{\mu }})$ is generally given by $\frac{1}{3!}\epsilon ^{\mu
\nu \rho }\partial _{\mu }n^{a}\partial _{\nu }n^{b}\partial _{\nu
}n^{b}\varepsilon _{abc};$ it maps the spacetime volume $d^{3}\mathbf{\xi }%
=dt\wedge dx\wedge dy$ into the target space volume $d^{3}\mathbf{n}%
=dn^{1}\wedge dn^{2}\wedge dn^{3}$. In this regards, recall that these two
3D volumes can be expressed in covariant manners by using the completely
antisymmetric tensors $\epsilon _{\mu \nu \rho }$ and $\varepsilon _{abc}$
introduced earlier; and as noticed before play a central role in topology.
The target space volume $d^{3}\mathbf{n}$ can be expressed like $\frac{1}{3!}%
\varepsilon _{abc}dn^{a}\wedge dn^{b}\wedge dn^{c}$; and a similar relation
can be also written down for the spacetime volume $d^{3}\mathbf{\xi }$.
Notice also that by substituting the differentials $dn^{a}$ by their
expansions $(\frac{\partial n^{a}}{\partial \xi ^{\mu }})d\xi ^{\mu }$; and
putting back into $d^{3}\mathbf{n},$ we obtain the relation $d^{3}\mathbf{n}=%
\mathfrak{J}_{{\small 3D}}d^{3}\mathbf{\xi }$ where $\mathfrak{J}_{{\small 3D%
}}$ is precisely the Jacobian $\det (\frac{\partial n^{a}}{\partial \xi
^{\mu }})$. $\left( 2\right) $ The conservation law $\partial _{\mu }(J^{\mu
})_{{\small 3D}}=0$ has a geometric origin; it follows from the field
constraint relation $n_{1}^{2}+n_{2}^{2}+n_{3}^{2}=1$ degenerating the
volume of the 3D target space down to a surface. This constraint relation
describes a unit 2-sphere $\mathbb{S}_{\left( \mathbf{n}\right) }^{2}$; and
so a vanishing volume $\left. d^{3}\mathbf{n}\right \vert _{\mathbb{S}%
_{\left( \mathbf{n}\right) }^{2}}=0;$ thus leading to $\mathfrak{J}_{3D}=0$
and then to the above continuity equation. Having the explicit expression (%
\ref{cs}) of the topological current $J^{\mu }$ in terms of the magnetic
texture field $n\left( \xi \right) $, we turn to determine the associated
topological charge $Q=\int dxdyJ^{0}$ with charge density $J^{0}$ given by $%
\frac{1}{8\pi }\varepsilon _{abc}\epsilon ^{0ij}\left( \partial
_{i}n^{b}\partial _{j}n^{c}\right) n^{a}$. Substituting $\epsilon
^{0ij}dx\wedge dy$ by $d\xi ^{i}\wedge d\xi ^{j}$, we have $J^{0}dx\wedge dy=%
\frac{1}{8\pi }\varepsilon _{abc}n^{a}\left( dn^{b}\wedge dn^{c}\right) $.
Moreover using the differentials $dn^{b}=u^{b}d\beta +v^{b}\sin \alpha
d\alpha $, we can calculate the area $dn^{b}\wedge dn^{c}$ in terms of the
angles $\alpha $ and $\beta ;$ we find $2n_{a}(\sin \alpha )d\beta \wedge
d\alpha $ where we have used $\varepsilon _{abc}\left(
u^{b}v^{c}-u^{c}v^{b}\right) =2n_{a}.$ So, the topological charge $Q$ reads
as $\frac{1}{4\pi }\int_{\mathbb{S}_{\mathbf{n}}^{2}}\left( \sin \beta
\right) d\alpha d\beta $ which is equal to 1. In fact this value is just the
unit charge; the general value is an integer $Q=N$ with N being the winding
number $\pi _{2}(\mathbb{S}_{\mathbf{n}}^{2});$ see below. Notice that $%
J^{0} $ can be also presented like
\begin{equation}
J^{0}=\frac{\varepsilon _{abc}}{8\pi }n^{a}\left( \frac{\partial n^{b}}{%
\partial x}\frac{\partial n^{c}}{\partial y}-\frac{\partial n^{b}}{\partial y%
}\frac{\partial n^{c}}{\partial x}\right)
\end{equation}%
Replacing $n_{a}$ by their expression in terms of the angles $\left( \sin
\beta \cos \alpha ,\sin \beta \sin \alpha ,\cos \beta \right) $, we can
bring the above charge density $J^{0}$ into two equivalent relations; first
into the form like $\frac{\sin \beta }{4\pi }\left( \frac{\partial \beta }{%
\partial x}\frac{\partial \alpha }{\partial y}-\frac{\partial \beta }{%
\partial y}\frac{\partial \beta }{\partial x}\right) $; and second as $\frac{%
1}{4\pi }\frac{\partial \left[ \alpha ,\cos \beta \right] }{\partial \left[
x,y\right] }$ which is nothing but the Jacobian of the transformation from
the $\left( x,y\right) $ space to the unit 2-sphere with angular variables $%
\left( \alpha ,\beta \right) $. The explicit expression of $\left(
n_{1},n_{2},n_{3}\right) $ in terms of the $\left( x,y\right) $ space
variables is given by
\begin{equation}
n_{1}=\frac{2x}{x^{2}+y^{2}+1}\quad ,\quad n_{2}=\frac{2y}{x^{2}+y^{2}+1}%
\quad ,\quad n_{3}=\frac{x^{2}+y^{2}-1}{x^{2}+y^{2}+1}
\end{equation}%
but this is nothing but the stereographic projection of the 2-sphere $%
\mathbb{S}_{\mathbf{\xi }}^{2}$ on the real plane. So, the field $n^{a}$
defines a mapping between $\mathbb{S}_{\mathbf{\xi }}^{2}$ towards $\mathbb{S%
}_{\mathbf{n}}^{2}$ with topological charge given by the winding number $%
\mathbb{S}_{\mathbf{n}}^{2}$ around $\mathbb{S}_{\mathbf{n}}^{2}$; this
corresponds just to the homotopy property $\pi _{2}(\mathbb{S}_{\mathbf{n}%
}^{2})=N$.

\section{Three dimensional magnetic skyrmions}

In this section, we study the \textrm{dynamics} of the 3d skyrmion and its
topological properties both in target space $\mathbb{R}_{\boldsymbol{n}}^{4}$
(with euclidian metric $\delta _{AB}$) and in 4D spacetime $\mathbb{R}_{\xi
}^{1,3}$ parameterised by $\xi ^{\mu }=\left( t,x,y,z\right) $ (with
Lorentzian metric $g_{\mu \nu }$). The spacetime dynamics of the 3d skyrmion
is described by a four component field $\boldsymbol{n}_{A}\left( \xi \right)
$ obeying a constraint relation $f\left( \boldsymbol{n}\right) =1$; here the
$f\left( \boldsymbol{n}\right) $ is given by the quadratic form $\boldsymbol{%
n}_{A}\boldsymbol{n}^{A}$ invariant under SO$\left( 4\right) $
transformations isomorphic to SU$\left( 2\right) \times $SU$\left( 2\right) $%
. The structure of the topological current of the 3d skyrmion is encoded in
two types of Levi-Civita tensors namely the target space $\varepsilon
_{ABCD} $ and the spacetime $\epsilon _{\mu \nu \rho \tau }$ extending their
homologue concerning the kinks and 2d skyrmions.

\subsection{From 2d skyrmion to 3d homologue}

As for the 1d and 2d solitons considered previous section, the spacetime
dynamics of the 3d skyrmion in $\mathbb{R}^{1,3}$ is described by a field
action $\mathcal{S}_{4D}=\int dtL_{4D}$ with Lagrangian realized as the
space integral $\int dxdydz\mathcal{L}_{4D}$. Generally, the Lagrangian
density $\mathcal{L}_{4D}$ is a function of the soliton $\boldsymbol{n}%
\left( t,x,y,z\right) $ which is a real 4-component field [$\boldsymbol{n}%
=\left( n_{1},n_{2},n_{3},n_{4}\right) $] constrained like $f\left[
\boldsymbol{n}\left( \xi \right) \right] =1.$ For self ineracting field, the
typical field expression of $\mathcal{L}_{4D}$ is given by $-\frac{1}{2}%
\left( \partial _{\mu }\boldsymbol{n}\right) ^{2}-V\left( \boldsymbol{n}%
\right) -\Lambda \left[ f\left( \boldsymbol{n}\right) -1\right] $ where $%
V\left( \boldsymbol{n}\right) $ is a scalar potential; and where the
auxiliary field $\Lambda \left( \xi \right) $ is a Lagrange multiplier
carrying the field constraint. This density $\mathcal{L}_{4D}$ reads in
terms of the Hamiltonian as $\mathbf{\Pi .}\frac{\mathbf{\partial n}}{%
\partial t}-\mathcal{H}\left( \mathbf{n}\right) .$ Below, we consider a
4-component skyrmionic field constrained as $\boldsymbol{n}\mathbf{.}%
\boldsymbol{n}=1$; and focuss on a simple Lagrangian density $\mathcal{L}%
_{\circ }=-\frac{1}{2}\left( \partial _{\mu }\boldsymbol{n}\right) \left(
\partial ^{\mu }\boldsymbol{n}\right) -\Lambda \left[ \boldsymbol{n}\mathbf{.%
}\boldsymbol{n}-1\right] $ to describe the degrees of freedom of $%
\boldsymbol{n}$. Being a unit 4-component vector, we can solve the
constraint $\boldsymbol{n}\mathbf{.}\boldsymbol{n}=1$ in terms of three
angular angles $\left( \alpha ,\beta ,\gamma \right) ;$ by setting%
\begin{equation}
\boldsymbol{n}=\left( \mathbf{m}\sin \gamma ,\cos \gamma \right) \quad
,\quad \mathbf{m}=\left( \sin \beta \cos \alpha ,\sin \beta \sin \alpha
,\cos \beta \right)  \label{nm}
\end{equation}%
where $\mathbf{m}$ is a unit 3-vector parameterising the unit sphere $%
\mathbb{S}_{\left[ \mathbf{\alpha }\right] }^{2}$. Putting this field
realisation back into $\mathcal{L}_{\circ }$, we obtain $\frac{-\cos 2\gamma
}{2}\left( \partial _{\mu }\gamma \right) ^{2}-\frac{1-\cos 2\gamma }{4}%
\left( \partial _{\mu }\mathbf{m}\right) ^{2}-\Lambda \left[ \mathbf{m.m}-1%
\right] .$ Notice that by restricting the 4D spacetime $\mathbb{R}^{1,3}$ to
the 3D hyperplane z=const; and by fixing the component field $\gamma $ to $%
\frac{\pi }{2}$, the above Lagrangian density reduces to the one describing
the spacetime dynamics of the 2d skyrmion. Notice also that we can expand
the differential $d\boldsymbol{n}_{A}$ in terms of $d\gamma ,d\beta ,d\alpha
$; we find the following%
\begin{equation}
d\boldsymbol{n}_{a}=m_{a}\cos \gamma d\gamma +\sin \gamma \left( u_{a}d\beta
+v_{a}\sin \beta d\alpha \right) \quad ,\quad d\boldsymbol{n}_{4}=-\sin
\gamma d\gamma
\end{equation}%
For convenience, we sometimes refer to the three $\left( \alpha ,\beta
,\gamma \right) $ collectively like $\alpha _{a}=\left( \alpha _{1},\alpha
_{2},\alpha _{3}\right) ;$ so we have $d\boldsymbol{n}^{A}=E_{a}^{A}d\alpha
^{b}$ with $E_{a}^{A}=\frac{\partial \boldsymbol{n}^{A}}{\partial \mathrm{%
\alpha }^{a}}.$

\subsection{Conserved topological current}

First, we investigate the topological properties of the 3d skyrmion from the
target space view; that is without using the spacetine variables $\left(
t,x,y,z\right) =\xi ^{\mu }$. Then, we turn to study the induced topological
properties of the 3d skyrmion viewed from the side of the 4D space time $%
\mathbb{R}^{1,3}$.

\subsubsection{Topological current in target space}

The 3d skyrmion field is described by a real four component vector $%
\boldsymbol{n}_{A}$ subject to the constraint relation $\boldsymbol{n}_{A}%
\boldsymbol{n}^{A}=1$; so the soliton has SO$\left( 4\right) \sim SO\left(
3\right) _{1}\times SO\left( 3\right) _{2}$ symmetry leaving invariant the
condition $\boldsymbol{n}_{A}\boldsymbol{n}^{A}=1$ that reads explicitly as $%
\left( n_{1}\right) ^{2}+\left( n_{2}\right) ^{2}+\left( n_{3}\right)
^{2}+\left( n_{4}\right) ^{2}=1.$ The algebraic condition $f\left[
\boldsymbol{n}\right] =1$ induces in turns the constraint equation $df=0$
leading to $\boldsymbol{n}^{A}d\boldsymbol{n}_{A}=0$ and showing that $%
\boldsymbol{n}_{A}$ and $d\boldsymbol{n}_{A}$ orthogonal 4-vectors in $%
\mathbb{R}_{\left( \boldsymbol{n}\right) }^{4}$. From this constraint, we
can construct $\left( \boldsymbol{n}^{A}d\boldsymbol{n}^{B}-\boldsymbol{n}%
^{B}d\boldsymbol{n}^{A}\right) /2$ which is a 4$\times $4 antisymmetric
matrix $\Omega ^{\left[ AB\right] }$ \textrm{generating} the SO$\left(
4\right) $ rotations; this $\Omega ^{\left[ AB\right] }$ contains 3+3
degrees of freedom generating the two $SO\left( 3\right) _{1}$ and $SO\left(
3\right) _{2}$ making SO$\left( 4\right) $; the first three degrres are
given by $\Omega ^{\left[ ab\right] }$ with $a,b=1,2,3$; and the other three
concern $\Omega ^{\left[ a4\right] }.$ Notice also that, from the view of
the target space, the algebraic relation $\boldsymbol{n}_{A}\boldsymbol{n}%
^{A}=1$ describes a unit 3-sphere $\mathbb{S}_{\boldsymbol{n}}^{3}$ sitting
in $\mathbb{R}_{\boldsymbol{n}}^{4}$; as such its volume 4-form $d^{4}%
\boldsymbol{n}$, which reads as $\frac{1}{4!}\varepsilon _{ABCD}d\boldsymbol{%
n}^{A}\wedge d\boldsymbol{n}^{B}\wedge d\boldsymbol{n}^{C}\wedge d%
\boldsymbol{n}^{D},$ vanishes identically when restricted to the 3-sphere;
i.e: $\left. d^{4}\boldsymbol{n}\right \vert _{\mathbb{S}_{\boldsymbol{n}%
}^{3}}=0$. This vanishing property of $d^{4}\boldsymbol{n}$ on $\mathbb{S}_{%
\boldsymbol{n}}^{3}$ is a key ingredient in the derivation of the
topological current $\boldsymbol{J}$ of the 3D skyrmion and its conservation
$d\boldsymbol{J}=0$. Indeed, because of the property $d^{2}=0$ (where we
have hidden the wedge product $\wedge $), it follows that $d^{4}\boldsymbol{n%
}$ can be expressed as $d\boldsymbol{J}$ with the 3-form $\boldsymbol{J}$
given by
\begin{equation}
\boldsymbol{J}=\frac{1}{4!}\varepsilon _{ABCD}\boldsymbol{n}^{A}d\boldsymbol{%
n}^{B}d\boldsymbol{n}^{C}d\boldsymbol{n}^{D}
\end{equation}%
This 3-form describes precisely the topological current in the target space;
this is because on $\mathbb{S}_{\boldsymbol{n}}^{3}$, the 4-form $d^{4}%
\boldsymbol{n}$ vanishes; and then $d\boldsymbol{J}$ vanishes. By solving,
the skyrmion field constraint $\boldsymbol{n}_{A}\boldsymbol{n}^{A}=1$ in
terms of three angles $\mathrm{\alpha }_{a}$ as given by Eq(\ref{nm}); with
these angular coordinates, we have mapping $\boldsymbol{f}:\mathbb{R}_{%
\boldsymbol{n}}^{4}\rightarrow \mathbb{S}_{\boldsymbol{n}}^{3}$ with $%
\mathbb{S}_{\boldsymbol{n}}^{3}\simeq \mathbb{S}_{\mathbf{\alpha }}^{3}$. By
expanding the differentials like $d\boldsymbol{n}^{A}=E_{a}^{A}d\mathrm{%
\alpha }^{a}$ with $E_{a}^{A}=\frac{\partial \boldsymbol{n}^{A}}{\partial
\mathrm{\alpha }^{a}}$; then the conserved current on the 3-sphere $\mathbb{S%
}_{\mathbf{\alpha }}^{3}$ reads as follows
\begin{equation}
\boldsymbol{J}=\frac{1}{4!3!}\varepsilon _{ABCD}(\boldsymbol{n}%
^{A}E_{b}^{B}E_{c}^{C}E_{d}^{D})\varepsilon ^{abc}d^{3}\mathbf{\alpha }
\end{equation}%
where we have substituted the 3-form $d\mathrm{\alpha }^{b}d\mathrm{\alpha }%
^{c}d\mathrm{\alpha }^{d}$ on the 3-sphere $\mathbb{S}_{\mathbf{\alpha }%
}^{3} $ by the volume 3-form $\varepsilon ^{abc}d^{3}\mathbf{\alpha }$. In
this regards, recall that the volume of the 3-sphere is $\int_{\mathbb{S}%
_{\left( \boldsymbol{n}\right) }^{3}}d^{3}\mathbf{\alpha }=\frac{\pi ^{2}}{2}
$.

\subsubsection{Topological symmetry in spacetime}

In the spacetime $\mathbb{R}^{1,3}$ with coordinates $\xi ^{\mu }=\left(
t,x,y,z\right) $, the 3d skyrmion is described by a four component field $%
\boldsymbol{n}_{A}\left( \xi \right) $ and is subject to the local
constraint relation $\boldsymbol{n}_{A}\boldsymbol{n}^{A}=1$. A typical
\textrm{static} configuration of the 3d skyrmion is obtained by solving the
field sonctraint in terms of the space coordinates; it is given by Eq(\ref%
{nm}) with the local space time fields $\mathbf{m}\left( \xi \right) $ and $%
\gamma \left( \xi \right) $ thought of as follows%
\begin{equation}
\mathbf{m}\left( \xi \right) =\left( \frac{x}{r},\frac{y}{r},\frac{z}{r}%
\right) \quad ,\quad \gamma \left( \xi \right) =\arcsin \frac{2rR}{%
r^{2}+R^{2}}=\arccos \frac{r^{2}-R^{2}}{r^{2}+R^{2}}  \label{mg}
\end{equation}%
with $r=\sqrt{x^{2}+y^{2}+z^{2}},$ giving the radius of $\mathbb{S}_{\xi
}^{2},$ and $R$ associated with the circle $\mathbb{S}_{\xi }^{1}$ fibered
over $\mathbb{S}_{\xi }^{2}$; the value $R=r$ corresponds to $\gamma =\frac{%
\pi }{2}$ and $R>>r$ to $\gamma =\pi $. Notice that $\gamma \left( \xi
\right) $ in eq(\ref{mg}) has a spherical symmetry as it is a function only
of $r$ (no angles $\alpha ,\beta ,\gamma $). Moreover, as this configuration
obeys $\sin \gamma \left( 0\right) =0$ and $\sin \gamma \left( \infty
\right) =0$; we assume $\gamma \left( 0\right) =n_{0}\pi $ and $\gamma
\left( \infty \right) =n_{\infty }\pi $. Putting these relations back into (%
\ref{nm}), we obtain the following configuration%
\begin{equation}
\boldsymbol{\tilde{n}}=\left( \frac{2xR}{r^{2}+R^{2}},\frac{2yR}{r^{2}+R^{2}}%
,\frac{2zR}{r^{2}+R^{2}},\frac{r^{2}-R^{2}}{r^{2}+R^{2}}\right)
\end{equation}%
describing a compactification of the space $\mathbb{R}_{\xi }^{3}$ into $%
\mathbb{S}_{\xi }^{3}$ which is homotopic to $\mathbb{S}_{\boldsymbol{n}%
}^{3} $. From this view, the $\boldsymbol{\tilde{n}}:\xi \rightarrow
\boldsymbol{\tilde{n}}\left( \xi \right) $ is then a mapping from $\mathbb{S}%
_{\xi }^{3}$ into $\mathbb{S}_{\boldsymbol{n}}^{3}$ with topological charge
given by the winding number characterising the wrapping $\mathbb{S}_{%
\boldsymbol{n}}^{3}$ on $\mathbb{S}_{\xi }^{3}$; and for which we have the
property $\pi _{3}(\mathbb{S}_{n}^{3})=\mathbb{Z}$. In this regards, recall
that 3-spheres $\mathbb{S}^{3}$ have a Hopf fibration given by a circle $%
\mathbb{S}^{1}$ sitting over $\mathbb{S}^{2}$ (for short $\mathbb{S}^{3}\sim
\mathbb{S}^{1}\ltimes \mathbb{S}^{2}$); this non trivial fibration can be
viewed from the relation $\mathbb{S}^{3}\sim SU\left( 2\right) $ and the
factorisation $U\left( 1\right) \times SU\left( 2\right) /U\left( 1\right) $
with the coset $SU\left( 2\right) /U\left( 1\right) $ identified with $%
\mathbb{S}^{2}$; and $U\left( 1\right) $ with $\mathbb{S}^{1}$. Applying
this fibration to $\mathbb{S}_{\boldsymbol{\xi }}^{3}$ and $\mathbb{S}_{%
\boldsymbol{n}}^{3}$, it follows that $\boldsymbol{\tilde{n}}:\mathbb{S}%
_{\xi }^{3}\rightarrow \mathbb{S}_{\boldsymbol{n}}^{3};$ and the same thing
for the bases $\mathbb{S}_{\xi }^{2}\rightarrow \mathbb{S}_{\boldsymbol{n}%
}^{2}$ and for the fibers $\mathbb{S}_{\xi }^{1}\rightarrow \mathbb{S}_{%
\boldsymbol{n}}^{1}$. Returning to the topological current and the conserved
topological charge $Q=\int_{\mathbb{R}^{3}}d^{3}\mathbf{r}J^{0}\left( t,%
\mathbf{r}\right) $, notice that in space time the differential $d%
\boldsymbol{n}^{A}$ expands like $\left( \partial _{\mu }\boldsymbol{n}%
^{A}\right) d\xi ^{\mu }$; then using the duality relation $J_{\left[ \nu
\rho \tau \right] }=\epsilon _{\mu \nu \rho \tau }J^{\mu }$, we find, up to
a normalisation by the volume of the 3-sphere $\pi ^{2}/2$, the expression
of the topological current $J^{\mu }\left( \xi \right) $ in terms of the 3D
skyrmion field%
\begin{equation}
J^{\mu }=\frac{1}{12\pi ^{2}}\epsilon ^{\mu \nu \rho \tau }n^{a}\partial
_{\nu }n^{b}\partial _{\rho }n^{c}\partial _{\tau }n^{c}\varepsilon _{abcd}
\end{equation}%
In terms of the angular variables, this current reads like $\mathcal{N}%
\partial _{\nu }\alpha \partial _{\rho }\beta \partial _{\tau }\gamma
\epsilon ^{\mu \nu \rho \tau }$ with $\mathcal{N}=\frac{1}{2\pi ^{2}}\left(
\sin \beta \right) \left( \sin \gamma \right) ^{2}$. From this current
expression, we can determine the associated topological charge $Q$ by space
integration over the charge density
\begin{equation}
J^{0}\left( t,\mathbf{r}\right) =-\frac{\sin ^{2}\gamma }{2\pi ^{2}r^{2}}%
\frac{d\gamma }{dr}
\end{equation}%
Because of its spherical symmetry, the space volume $d^{3}\mathbf{r}$ can be
substituted by $4\pi r^{2}dr$; then the charge $Q$ reads as the integral $-%
\frac{4\pi }{2\pi ^{2}}\int_{\gamma \left( 0\right) }^{\gamma \left( \infty
\right) }\sin ^{2}\gamma d\gamma $ whose integration leads to the sum of two
terms coming from the integration of $\sin ^{2}\gamma =\frac{1}{2}-\frac{1}{2%
}\cos 2\gamma $. The integral first reads as $\frac{1}{\pi }\left[ \gamma
\left( 0\right) -\gamma \left( \infty \right) \right] $; by substituting $%
\gamma \left( 0\right) =n_{0}\pi $, it contributes like $N\pi $. The
integral of the second tem gives $\frac{1}{2\pi }\left[ \sin 2\gamma \left(
0\right) -\sin 2\gamma \left( \infty \right) \right] $; it vanishes
identically. So the topological charge is given by
\begin{equation}
Q=\frac{\gamma \left( 0\right) -\gamma \left( \infty \right) }{\pi }=N
\end{equation}

\section{Effective dynamics of skyrmions}

In this section, we investigate the effective dynamics of a point- like
skyrmion in a ferromagnetic background field while focussing on the 2d
configuration. First, we derive the effective equation of a rigid skyrmion
and comment on the underlying effective Lagrangian. We also describe the
similarity with the dynamics of an electron in a background electromagnetic
field. Then, we study the effect of dissipation on the skyrmion dynamics.

\subsection{Equation of a rigid skyrmion}

To get the effective equation of motion of a rigid skyrmion, we start by the
spin $\left( 0+1\right) D$ action $\mathcal{S}_{spin}=\int dtL_{spin}$
describing the time evolution of a coherent spin vector modeled by a
rotating magnetic moment $\mathbf{n}\left( t\right) $ with velocity $\mathbf{%
\dot{n}}=\frac{d\mathbf{n}}{dt}$; and make some accommodations. For that,
recall that the Lagrangian $L_{spin}$ has the structure $L_{B}-\hbar S%
\mathrm{\gamma }H$ where $L_{B}$ is the Berry term having the form $%
L_{B}=q_{e}\mathbf{A}.\mathbf{\dot{n}}$ with geometric (Berry) potential $%
\mathbf{A}\sim \left \langle \mathbf{n}|\mathbf{\dot{n}}\right \rangle $;
and where $H$ is the Hamiltonian of the magnetic moment $\mathbf{n}\left(
t\right) $ obeying the constraint $\mathbf{n}^{2}=1$. This magnetisation
constraint is solved by two free angles $\alpha \left( t\right) ,\beta
\left( t\right) ;$ they appear in the Berry term $L_{B}=-\hbar S\left(
1-\cos \beta \right) \frac{d\alpha }{dt}.$ Below, we think of the above
magnetisation as a ferromagnetic background $\mathbf{n}\left( \mathbf{r}%
\right) $ filling the spatial region of $\mathbb{R}_{\xi }^{1,d}$ with
coordinates $\xi =\left( t,\mathbf{r}\right) $; and of the skymion as a
massive point- like particle $\mathbf{R}\left( t\right) $ moving in this
background.

\subsubsection{Rigid skyrmion}

We begin by introducing the variables describing the skyrmion in the
magnetic background field $\mathbf{n}\left( \mathbf{r}\right) $. We denote
by $M_{s}$ the mass of the skyrmion, and by $\mathbf{R}$ and $\mathbf{\dot{R}%
}$ its space position and its velocity. For concreteness, we restrict the
investigation to the spacetime $\mathbb{R}_{\xi }^{1,2}$ and refer to $%
\mathbf{R}$ by the components $X_{i}=\left( X,Y\right) $ and to $\mathbf{r}$
by the components $x_{i}=\left( x,y\right) $. Because of the Euclidean
metric $\delta _{ij}$; we often we use both notations $X^{i}$ and $%
X_{i}=\delta _{ij}X^{j}$ without referring to $\delta _{ij}$. Furthermore;
we limit the discussion to the interesting case where the only source of
displacements in $\mathbb{R}_{\xi }^{1,2}$ is due to the skyrmion $\mathbf{R}%
\left( t\right) $ (rigid skyrmion). In this picture, the description of the
skyrmion $\mathbf{R}\left( t\right) $ dissolved in the background magnet $%
\mathbf{n}\left( \mathbf{r}\right) $ is given by%
\begin{equation}
\mathbf{n}\left( \mathbf{r},t\right) =\mathbf{n}\left[ \mathbf{r}-\mathbf{R}%
\left( t\right) \right]  \label{R}
\end{equation}%
In this representation, the velocity $\mathbf{\dot{n}}$ of the skyrmion
dissolved in the background magnet can be expressed into manners; either
like $-\dot{X}^{i}\frac{\partial \mathbf{n}}{\partial X^{i}}$; or as $\dot{X}%
^{i}\frac{\partial \mathbf{n}}{\partial x^{i}}$; this is because $\frac{%
\partial }{\partial X^{i}}=-\frac{\partial }{\partial x^{i}}$. With this
parametrisation, the dynamics of the skyrmion is described by an action $%
\mathcal{S}_{s}=\int dtL_{s}$ with Lagrangian given by a space integral $%
L_{s}=\frac{\hbar s}{\text{\texttt{a}}^{2}}\int d^{2}\mathbf{r}\mathcal{L}%
_{s}$ and spacetime density as follows%
\begin{equation}
\mathcal{L}_{s}=\mathrm{\gamma }\hbar S\mathcal{H}-\hbar S\mathcal{L}_{B}
\label{lh}
\end{equation}%
In this relation, the density $\mathcal{L}_{B}=$ $-\left( 1-\cos \beta
\right) \frac{\partial \alpha }{\partial t}$ where now the angular variables
are spacetime fields $\beta \left( t,\mathbf{r}\right) $ and $\alpha \left(
t,\mathbf{r}\right) $. Similarly, the density $\mathcal{H}$ is the
Hamiltonian density with arguments as $\mathcal{H}\left[ \mathbf{n},\partial
_{\mu }\mathbf{n},\mathbf{r}\right] $ and magnetic $\mathbf{n}$ as in Eq(\ref%
{R}). In this field action $\mathcal{S}_{s}$, the prefactor \texttt{a}$^{-2}$
is required by the continuum limit of lattice Hamiltonians $H_{latt}$ living
on a square lattice with spacing parameter \texttt{a}. Recall that for these
$H_{latt}$'s, one generally has discrete sums like $\sum_{\mu }\left(
...\right) $, $\sum_{\mu ,\nu }\left( ...\right) $ and so on; in the limit
where \texttt{a} is too small, these sums turn into 2D space integrals like
\texttt{a}$^{-2}\int d^{2}\mathbf{r}\left( ...\right) $. To fix ideas, we
illustrate this limit on the typical hamiltonian $H_{HDMZ}$, it describes
the Heisenberg model on the lattice $\mathbb{Z}^{2}$ augmented by the
Dzyaloshinskii-Moriya and the Zeeman interactions \textrm{\cite{1c,1c1}}
\begin{equation}
H_{HDMZ}=-J\sum_{\left \langle \mu ,\nu \right \rangle }\mathbf{n}\left(
\mathbf{r}_{\mu }\right) \mathbf{n}\left( \mathbf{r}_{\nu }\right)
-D\sum_{\mu ,\nu ,\rho }\mathbf{d}_{\mu }.\left[ \mathbf{n}\left( \mathbf{r}%
_{\nu }\right) \wedge \mathbf{n}\left( \mathbf{r}_{\rho }\right) \right]
\epsilon ^{\mu \nu \rho }-\sum_{\mu }\mathbf{B.n}\left( \mathbf{r}_{\mu
}\right)
\end{equation}%
with $\mathbf{r}_{\nu }=\mathbf{r}_{\mu }+\mathtt{a}\mathbf{e}_{\nu \mu }$;
that is $\mathbf{e}_{\nu \mu }=(\mathbf{r}_{\nu }-\mathbf{r}_{\mu })/\mathtt{%
a}$ where $\mathtt{a}$ is the square lattice parameter. So, the continuum
limit $\mathcal{H}$ of this lattice Hamiltonian involves the target space
metric $\delta _{ab}$ and the topological Levi-Civita tensor $\varepsilon
_{abc}$ of the target space $\mathbb{R}_{\mathbf{n}}^{3}$; it involves as
well the metric $g_{\mu \nu }$ and the completely antisymmetry $\epsilon
_{\mu \nu \rho }$ of the space time $\mathbb{R}_{\xi }^{1,2}$. In terms of $%
\delta _{ab}$ and $\varepsilon _{abc}$ tensors, the continuous hamiltonian
density reads as follows%
\begin{equation}
\mathcal{H}=\frac{J\mathtt{a}^{2}}{2}\delta _{ab}\partial ^{i}n^{a}\partial
_{i}n^{b}+\mathtt{a}\varepsilon _{abc}d_{\mu }^{a}\left( n^{b}D_{\nu \rho
}n^{c}\right) \epsilon ^{\mu \nu \rho }-\mathbf{B}.\mathbf{n}
\end{equation}%
with $\nabla _{\nu \rho }=\mathbf{e}_{\nu \mu }.\mathbf{\nabla }$. Below, we
set J=1 and, to factorise out the normalisation factor $\mathtt{a}^{2}$,
scale the parameters of the model like $d_{\mu }^{a}=\mathtt{a}\tilde{d}%
_{\mu }^{a}$ and $\mathbf{B}=\mathtt{a}^{2}\mathbf{\tilde{B}}$. For
simplicity, we sometimes set as well \texttt{a}$=1.$

\subsubsection{Skyrmion equation without dissipation}

To get the effective field equation of motion of the point- like skyrmion
without dissipation, we calculate the vanishing condition of the functional
variation of the action; that is $\delta (\int dtd^{2}\mathbf{r}\mathcal{L)}%
=0$. General arguments indicate that the effective equation of the skyrmion
with topological charge $q_{s}$ in the background magnet has the form
\begin{equation}
M_{s}\ddot{X}_{i}=q_{s}\mathcal{E}_{i}+q_{s}\epsilon _{ijz}\dot{X}^{j}%
\mathcal{B}^{z}  \label{eq}
\end{equation}%
from which one can wonder the effective Lagrangian describing the effective
dynamics of the skyrmion. It is given by
\begin{equation}
L_{s}=\frac{M_{s}}{2}\delta _{ij}\dot{X}^{i}\dot{X}^{j}-q_{s}\epsilon _{zij}%
\mathcal{B}^{z}X^{i}\dot{X}^{j}-q_{s}\mathcal{V}\left( X\right)  \label{qe}
\end{equation}%
Notice that the right hand of equation (\ref{eq}) looks like the usual
Lorentz force ($q_{e}\mathbf{E}+q_{e}\mathbf{\dot{r}\wedge B}$) of a moving
electron with $q_{e}$ in an external electromagnetic field $\left(
E_{i},B_{i}\right) ;$ the corresponding Lagrangian is $\frac{m}{2}\mathbf{%
\dot{r}}^{2}+q_{e}\mathbf{B.}(\mathbf{r\wedge \dot{r})-}q_{e}\mathbf{E.r}$.
This similarity between the skyrmion and the electron in background fields
is because the skyrmion has a topological charge $q_{s}$ that can be put in
correspondence with $q_{e}$; and, in the same way, the background field
magnet $\mathcal{E}_{i},\mathcal{B}_{i}$ can be also put in correspondence
with the electromagnetic field $\left( E_{i},B_{i}\right) $. To rigourously
derive the spacetime eqs(\ref{eq}-\ref{qe}), we need to perform some
manipulations relying on computing the effective expression of $\mathcal{S}%
_{s}=\hbar S\int dt(\int d^{2}\mathbf{r}\mathcal{L}_{s})$ and its \textrm{%
time} variation $\delta \mathcal{S}_{s}=0$. However, as $\mathcal{L}_{s}$
has two terms like $\mathrm{\gamma }\hbar S\mathcal{H}-\hbar S\mathcal{L}%
_{B} $, the calculations can be split in two stages; the first stage
concerns the block $\mathrm{\gamma }\hbar S\int d^{2}\mathbf{r}\mathcal{H}$
with $\mathcal{H}\left[ \mathbf{n},\partial _{\mu }\mathbf{n},\mathbf{r}%
\right] $ which is a function of the magnetic texture (\ref{R}); that is $%
\mathbf{n}\left( \mathbf{r}-\mathbf{R}\right) $. The second stage regards
the determination of the integral $\hbar S\int d^{2}\mathbf{r}\mathcal{L}%
_{B} $. The computation of the first term is straightforwardly identified;
by performing a space shift $\mathbf{r}\rightarrow \mathbf{r}+\mathbf{R}$,
the Hamiltonian density becomes $\mathcal{H}\left[ \mathbf{n},\partial _{\mu
}\mathbf{n},\mathbf{r}+\mathbf{R}\right] $ with $\mathbf{n}\left( \mathbf{r}%
\right) $ and where the dependence in $\mathbf{R}$ becomes explicit; thus
allowing to think of the integral $\mathrm{\gamma }\hbar S\int d^{2}\mathbf{r%
}\mathcal{H}$ as nothing but the scalar energy potential $\mathcal{V}\left(
\mathbf{R}\right) =\hbar S\mathrm{\gamma }\int d^{2}\mathbf{r}\mathcal{H}%
\left( t,\mathbf{r},\mathbf{R}\right) .$ So, we have%
\begin{equation}
\frac{\delta }{\delta X^{a}}\left( \hbar S\mathrm{\gamma }\int d^{2}\mathbf{r%
}\mathcal{H}\right) =\frac{\partial \mathcal{V}}{\partial X^{a}}
\end{equation}%
Concerning the calculation of the $\hbar S\int d^{2}\mathbf{r}\mathcal{L}%
_{B} $, the situation is somehow subtle; we do it in two steps; first we
calculate the $(\delta \int d^{2}\mathbf{r}\mathcal{L}_{B})$ because we know
the variation $\frac{\delta \mathcal{L}_{B}}{\delta n^{a}}$ which is equal
to $\frac{1}{2}\varepsilon _{abc}n^{b}\dot{n}^{c}$. Then, we turn backward
to determine $\hbar S\int d^{2}\mathbf{r}\mathcal{L}_{B}$ by integration. To
that purpose, recall also that the Berry term $\mathcal{L}_{B}$ is given by $%
-\left( 1-\cos \beta \right) \frac{\partial \alpha }{\partial t}$; and its
variation $\frac{\delta \mathcal{L}_{B}}{\delta n^{a}}\frac{\partial n^{a}}{%
\partial X^{j}}\partial X^{j}$ is equal to $\frac{1}{2}\varepsilon
_{abc}n^{b}\dot{n}^{c}$. To determine the time variation $\delta
L_{B}=\delta \int d^{2}\mathbf{r}\delta \mathcal{L}_{B}$, we first expand it
like $\int d^{2}\mathbf{r}\frac{\delta \mathcal{L}}{\delta n^{a}}\delta
n^{a} $; and use $\delta n^{a}=-\frac{\partial n^{a}}{\partial X^{j}}%
\partial X^{j} $ to put it into the form -$\int d^{2}\mathbf{r}\frac{\delta
\mathcal{L}_{B}}{\delta n^{a}}\frac{\partial n^{a}}{\partial X^{j}}\partial
X^{j}$. Then, substituting $\frac{\delta \mathcal{L}_{B}}{\delta n^{a}}$ by
its expression $\frac{1}{2}\varepsilon _{abc}n^{b}\dot{n}^{c}$ with $\dot{n}%
^{c}$ expanded like -$\frac{\partial n^{c}}{\partial X^{i}}\dot{X}^{i}$, we
end up with%
\begin{equation}
\delta L_{B}=2\hbar S\left( \int d^{2}\mathbf{r}\frac{1}{2}\varepsilon
_{abc}n^{b}\frac{\partial n^{c}}{\partial x^{i}}\frac{\partial n^{a}}{%
\partial x^{j}}\right) \epsilon ^{ij}\left( \dot{X}\delta Y-\dot{Y}\delta
X\right)
\end{equation}%
Next, using the relation $\epsilon ^{ij}d^{2}\mathbf{r}=dx^{i}\wedge dx^{j}$%
, the first factor becomes $\int \frac{1}{2}\varepsilon
_{abc}n^{a}dn^{b}\wedge dn^{c}$ gives precisely the \textrm{skyrmion
topological charge} $q_{s}.$ So, the resulting $\delta L_{B}$ reduces to $%
2q_{s}\hbar S\left( \dot{X}\delta Y-\dot{Y}\delta X\right) $ that reads also
like
\begin{equation}
\delta L_{B}=-2q_{s}\hbar S\epsilon _{ij}\dot{X}^{i}\delta X^{j}
\end{equation}%
This variation is very remarkable because it is contained in the variation
of the effective coupling $L_{B}^{int}=-2q_{s}\hbar S\epsilon _{ij}\dot{X}%
^{i}X^{j}$ which can be presented like $L_{B}^{int}=-q_{s}\mathcal{A}_{i}%
\dot{X}^{i}$\ where we have set $\mathcal{A}_{i}=2\hbar S\epsilon
_{zij}X^{j} $; this vector can be interpreted as the vector potential of an
effective magnetic field $\mathcal{B}^{z}=\frac{1}{2}\epsilon ^{zij}\partial
_{i}\mathcal{A}_{i}$. By adding the kinetic term $\frac{M_{s}}{2}\dot{X}^{i}%
\dot{X}_{i}$, we end up with an effective Lagrangian $L_{B}$ associated with
the Berry term; it reads as follows $L_{B}=\frac{M_{s}}{2}\delta _{ij}\dot{X}%
^{i}\dot{X}^{j}-q_{s}\mathcal{A}_{i}\dot{X}^{i}.$ So, the effective
Lagrangian $L_{eff}$ describing the rigid 2d skyrmion in a ferromagnet is
\begin{equation}
L_{eff}=\frac{M_{s}}{2}\mathbf{\dot{R}}^{2}-q_{s}\mathbf{A}.\mathbf{\dot{R}-}%
\mathcal{V}\left( R\right)  \label{L}
\end{equation}%
From this Lagrangian, we learn the equation of the motion of the rigid
skyrmion namely $M_{s}\ddot{X}_{j}=f_{j}+4q_{s}\hbar S\epsilon _{zij}\dot{X}%
^{i}$; for the limit $M_{s}=0$, it reduces to $\dot{X}^{i}=\frac{1}{%
4q_{s}\hbar S}\epsilon ^{zji}f_{j}$.

\subsection{Implementing dissipation}

So far we have considered magnetic moment obeying the constraint $\mathbf{n}%
^{2}=1$ with time evolution given by the LL$\ $equation $\mathbf{\dot{n}}=-%
\mathrm{\gamma }\mathbf{f\wedge n}$ where the force $\mathbf{f}=-\frac{%
\mathbf{\partial }\mathcal{H}}{\partial \mathbf{n}}$. Using this equation,
we deduce the typical properties $\mathbf{n.\dot{n}}=\mathbf{f.\dot{n}}=0$
from which we learn that the time variation $\frac{dH}{dt}$ of the
Hamiltonian, which reads as $\hbar S\mathrm{\gamma }\int d^{2}\mathbf{r}%
\frac{\mathbf{\partial }\mathcal{H}}{\partial n^{a}}\dot{n}^{a}$, vanishes
identically as explicitly exhibited below,%
\begin{equation}
\frac{dH}{dt}=-\hbar S\mathrm{\gamma }\int d^{2}\mathbf{r}\left( \mathbf{f.%
\dot{n}}\right)  \label{fn}
\end{equation}%
In presence of dissipation, we loose energy; and so one expects that $\frac{%
dH}{dt}<0$; indicating that the rigid skyrmion has a damped dynamics. In
what follows, we study the effect of dissipation in the ferromagnet and
derive the damped skyrmion equation.

\subsubsection{Landau- Lifshitz- Gilbert equation}

Due to dissipation, the force $\mathbf{F}$ acting on the rigid skyrmion $%
\mathbf{R}\left( t\right) $ has two terms, the old conservative $\mathbf{f}=-%
\frac{\mathbf{\partial }\mathcal{H}}{\partial \mathbf{n}};$ and an extra
force $\delta \mathbf{f}$ linearly dependent in magnetisation velocity $%
\mathbf{\dot{n}}$. Due to this extra force $\delta \mathbf{f}=\mathbf{-}%
\frac{\mathrm{\alpha }}{\mathrm{\gamma }}\mathbf{\dot{n}}$, the LL\ equation
gets modified; its deformed expression is obtained by shifting the old force
$\mathbf{f}$ like $\mathbf{F=f-}\frac{\mathrm{\alpha }}{\mathrm{\gamma }}%
\mathbf{\dot{n}}$ with $\mathrm{\alpha }$ a positive damping parameter
(Gilbert parameter). As such, the previous LL relation gives the so called
Landau-Lifschitz-Gilbert (LLG) equation \textrm{\cite{2c,2c1}}
\begin{equation}
\mathbf{\dot{n}}=-\mathrm{\gamma }\mathbf{f\wedge n}+\mathrm{\alpha }\mathbf{%
\dot{n}\wedge n}  \label{LLG}
\end{equation}%
where its both sides have $\mathbf{\dot{n}}$. From this generalised
relation, we still have $\mathbf{n.\dot{n}}=0$ (ensuring $\mathbf{n}^{2}=1$%
); however $\mathbf{f.\dot{n}\neq 0}$ as it is equal to the Gilbert term
namely $-\mathrm{\alpha }\left( \mathbf{f\wedge n}\right) .\mathbf{\dot{n}}.$
Notice that eq(\ref{LLG}) still describe a rotating magnetic moment in the
target space ($d\mathbf{n}=0$); but with a different angular velocity $%
\mathbf{\Omega }$ which, in addition to $\mathbf{f}$, depends moreover on
the Gilbert parameter and the magnetisation $\mathbf{n}$. By factorising eq(%
\ref{LLG}) like $\mathbf{\dot{n}}=\mathbf{\Omega \wedge n},$ we find%
\begin{equation}
\mathbf{\Omega =}\frac{-\mathrm{\gamma }}{1+\mathrm{\alpha }^{2}}\left[
\mathbf{f}+\mathrm{\alpha }\mathbf{\left( \mathbf{f\wedge n}\right) }\right]
\end{equation}%
Notice that in presence of dissipation ($\mathrm{\alpha }\neq 0$), the
variation of the hamiltonian $\frac{dH}{dt}$ given by (\ref{fn}) is no
longer non vanishing; by first replacing $\mathbf{f.\dot{n}=}-\mathrm{\alpha
}\left( \mathbf{f\wedge n}\right) .\mathbf{\dot{n}}$ and putting back in it,
we get%
\begin{equation}
\frac{dH}{dt}=\mathrm{\alpha }\hbar S\int d^{2}\mathbf{r}\mathrm{\gamma }%
\left( \mathbf{f\wedge n}\right) .\mathbf{\dot{n}}
\end{equation}%
then, substituting $\mathrm{\gamma }\mathbf{f\wedge n}=\mathrm{\alpha }%
\mathbf{\dot{n}\wedge n}-\mathbf{\dot{n}}$; we find that $\frac{dH}{dt}$ is
given by $-\mathrm{\alpha }\hbar s\int d^{2}\mathbf{r}.\mathbf{\dot{n}}^{2}$
indicating that $\frac{dH}{dt}<0$; and consequently a decreasing energy $%
H\left( t\right) $ (loss of energy) while increasing time.

\subsubsection{Damped skyrmion equation}

To obtain the damped skyrmion equation due to the Gilbert term, we consider
the rigid magnetic moment $\mathbf{n}\left[ \mathbf{r}-\mathbf{R}\left(
t\right) \right] $; and compute the expression of the skyrmion velocity $%
\mathbf{\dot{R}}$ in terms of the conservative force $\mathbf{f}$ and the
parameter $\mathrm{\alpha }$. To that purpose we start from eq(\ref{LLG})
and multiply both equation sides by $\mathbf{\wedge }d\mathbf{n}$ while
assuming $\mathbf{f.n}=0$ (the conservative force transverse to
magnetisation), we get $\mathbf{\dot{n}\wedge }d\mathbf{n}=-\mathrm{\gamma }%
\left( \mathbf{f.}d\mathbf{n}\right) \mathbf{n}+\mathrm{\alpha }\left( d%
\mathbf{n.\dot{n}}\right) \mathbf{n}$. Then, multiply scalarly by $\mathbf{n,%
}$ which corresponds to a projection along the magnetisation, we obtain
\begin{equation}
\mathbf{n.}\left( \mathbf{\dot{n}\wedge }d\mathbf{n}\right) =-\mathrm{\gamma
}\left( \mathbf{f.}d\mathbf{n}\right) +\mathrm{\alpha }\left( d\mathbf{n.%
\dot{n}}\right)
\end{equation}%
\textrm{Substituting }$d\mathbf{n}$ and $\mathbf{\dot{n}}$ by their
expansions $dx^{i}\left( \partial _{i}\mathbf{n}\right) $ and $-\dot{X}%
^{i}\left( \partial _{i}\mathbf{n}\right) ,$ then multiplying by $\wedge
dx^{l}$; we end up with a relation involving $dx^{j}\wedge dx^{l}$ (which
reads as $\epsilon ^{zjl}d^{2}\mathbf{r}$); so we have
\begin{equation}
\dot{X}^{l}\left( 4\pi J^{0}d^{2}\mathbf{r}\right) =-\mathrm{\gamma }%
\epsilon ^{0lj}\left( \mathbf{f.}\partial _{j}\mathbf{n}\right) d^{2}\mathbf{%
r}+\mathrm{\alpha }\epsilon ^{0lj}\dot{X}_{j}\left( \partial _{j}\mathbf{n}%
\right) ^{2}d^{2}\mathbf{r}
\end{equation}%
where we have set $J^{0}=\frac{1}{2\pi }\epsilon ^{zij}\mathbf{n.}\left[
\left( \partial _{i}\mathbf{n}\right) \mathbf{\wedge }\left( \partial _{j}%
\mathbf{n}\right) \right] ,$ defining the magnetization density, and where
we have replaced $\left( \partial _{i}\mathbf{n.\partial }_{j}\mathbf{n}%
\right) $ by $\delta _{ij}\left( \partial _{k}\mathbf{n}\right) ^{2}$. By
integrating over the 2d space while using $\int J^{0}d^{2}\mathbf{r}=4\pi
q_{s}$ and setting $\eta _{j}=\frac{1}{4\pi }\int d^{2}\mathbf{r}\left(
\partial _{j}\mathbf{n}\right) ^{2}\equiv \mathrm{\eta }$, we arrive at the
relation%
\begin{equation}
4\pi q_{s}\dot{X}^{l}=\mathrm{\gamma }\epsilon ^{0lj}\int \left( \mathbf{f.}%
\partial _{j}\mathbf{n}\right) d^{2}\mathbf{r}-4\pi \mathrm{\eta \alpha }%
\epsilon ^{0lj}\dot{X}^{l}  \label{nf}
\end{equation}%
with $\epsilon ^{zxy}=-\epsilon _{zxy}=-1$. The remaining step is to replace
the conservative force $\mathbf{f}$\ by $-\frac{\partial \mathcal{H}}{%
\partial \mathbf{n}}$ and proceeds in performing the integral over $\left(
\mathbf{f.}\partial _{j}\mathbf{n}\right) $. Because of the explicit
dependence into $\mathbf{r}$, the $\mathbf{f.}\partial _{j}\mathbf{n}$ can
be expressed like $\partial _{j}^{\exp }\mathcal{H}-\partial _{j}^{tot}%
\mathcal{H}$; the explicit derivation term $\partial _{j}^{\exp }\mathcal{H}$
has been added because the Hamiltonian density has an explicit dependence $%
\mathcal{H}\left[ \mathbf{n},\partial _{\mu }\mathbf{n},\mathbf{r}\right] .$
Recall that $\partial _{j}^{tot}\mathcal{H}$ is given by $\partial
_{j}^{\exp }\mathcal{H}+\frac{\partial \mathcal{H}}{\partial \mathbf{n}}%
.\partial _{j}\mathbf{n}$ which is equal to $\partial _{j}^{\exp }\mathcal{H}%
-\mathbf{f}.\partial _{j}\mathbf{n.}$ Notice also that the term $\partial
_{j}^{\exp }\mathcal{H}$ can be also expressed like $-\frac{\partial
\mathcal{H}}{\partial \mathbf{R}}$. Therefore, the integral $\left( \mathbf{%
f.}\partial _{j}\mathbf{n}\right) d^{2}\mathbf{r}$ has two contributions
namely the $\int (\partial _{j}^{tot}\mathcal{H}\mathbf{)}d^{2}\mathbf{r}$
which, being a total derivative, vanishes identically; and the term $\int
(\partial _{j}^{\exp }\mathcal{H}\mathbf{)}d^{2}\mathbf{r}$ that gives $-%
\frac{\partial \mathcal{V}}{\partial \mathbf{R}}.$ Putting this value back
into (\ref{nf}), we end up with
\begin{equation}
4\pi q_{s}\dot{X}^{l}=\mathrm{\gamma }\epsilon ^{zlj}\left( \frac{\partial
\mathcal{V}}{\partial X^{j}}+\frac{4\pi \mathrm{\eta \alpha }}{\mathrm{%
\gamma }}\dot{X}^{l}\right)
\end{equation}%
Implementing the kinetic term of the skyrmion, we obtain the equation with
dissipation $M_{s}\mathbf{\ddot{R}}=-\frac{\partial \mathcal{V}}{\partial
\mathbf{R}}+G\left( \mathbf{z\wedge \dot{R}}-\frac{\mathrm{\eta \alpha }}{%
q_{s}}\mathbf{\dot{R}}\right) $ where the constant $G=\frac{4\pi \hbar Sq_{s}%
}{a^{2}}$ stands for the gyrostropic constant\textbf{.}

\section{Electron-skyrmion interaction}

In this section, we investigate the interacting dynamics between electrons
and skyrmions with spin transfer torque (STT) \textrm{\cite{3c}}. The
electron-skyrmion interaction is given by Hund coupling $J_{H}\mathbf{(}\Psi
^{\dagger }\sigma _{a}\Psi ).n^{a}$ which leads to emergent $SU\left(
2\right) $ gauge potential that mediate the interaction between the spin
texture $\mathbf{n}\left( t,\mathbf{r}\right) $ and the two spin states $%
(\Psi _{\uparrow },\Psi _{\downarrow })$ of the electron. We also study
other aspects of electron/skyrmion system like the limit of large Hund
coupling; and the derivation of the effective equation of motion of rigid
skyrmions with STT.

\subsection{Hund coupling}

We start by recalling that a magnetic atom (like iron, manganese, ...) can
be modeled by a localized magnetic moment $\mathbf{n}\left( t,\mathbf{r}%
\right) $ and mobile carriers represented by a two spin component field $%
\mathbf{\Psi }\left( t,\mathbf{r}\right) $; the components of the fields $%
\mathbf{n}$ and $\mathbf{\Psi }$ are respectively given by $n^{a}\left( t,%
\mathbf{r}\right) $ with a=1,2,3; and by $\Psi _{\alpha }\left( t,\mathbf{r}%
\right) $ with $\alpha =\uparrow \downarrow $. Using the electronic vector
density $\boldsymbol{j}_{{\small (e)}}=\Psi ^{\dagger }\mathbf{\sigma }\Psi $%
, the interaction between localised and itinerant electrons of the magnetic
atom are bound by the Hund coupling reading as $H_{e-n}=-J_{H}\mathbf{n.}%
\boldsymbol{j}_{{\small (e)}}$ with Hund parameter $J_{H}>0$ promoting
alignment of $\mathbf{n}$ and $\boldsymbol{j}_{{\small (e)}}$. So, the
dynamics of the interacting electron with the backround $\mathbf{n}$ is
given by the Lagrangian density $\mathcal{L}_{e}=\hbar \Psi ^{\dagger }\frac{%
i\partial }{\partial t}\Psi -H_{e-n}$ expanding as follows%
\begin{equation}
\mathcal{L}_{e}\left[ \Psi ,\mathbf{n}\right] =\hbar \Psi ^{\dagger }\frac{%
i\partial }{\partial t}\Psi -\Psi ^{\dagger }\left( \frac{\mathbf{P}^{2}}{2m}%
-J_{H}\mathbf{\sigma .n}\right) \Psi
\end{equation}%
where $\mathbf{P=}\frac{\hbar }{i}\nabla $ and $\mathbf{\sigma .n}=\sigma
^{x}n_{x}+\sigma ^{y}n_{y}+\sigma ^{z}n_{z}$.

\subsubsection{Emergent gauge potential}

Because of the ferromagnetic Hund coupling ($J_{H}>0$), the spin observable $%
\hat{S}_{e}^{z}=\frac{\hbar }{2}\sigma ^{z}$ of the conduction electron
tends to align with the orientation $\sigma ^{\mathbf{n}}=\mathbf{\sigma .n}$
of the magnetisation $\mathbf{n}$ --- with angle $\theta =(\widehat{\mathbf{e%
}_{z},\mathbf{n}})$---$;$ this alinement is accompanied by a local phase
change of the electronic wave function $\Psi $ which becomes $\mathbf{\psi }%
=U\mathbf{\Psi }$ where $U\left( t,\mathbf{r}\right) =e^{i\Theta \left( t,%
\mathbf{r}\right) }$ is a unitary $SU\left( 2\right) $ transformation
mapping $\sigma ^{z}$ into $\sigma ^{\mathbf{n}}$; that is $\sigma ^{\mathbf{%
n}}=U^{\dagger }\sigma ^{z}U$. For later use, we refer to the new two
components of the electronic field like $\mathrm{\psi }_{+\mathbf{n}},%
\mathrm{\psi }_{-\mathbf{n}}$ (for short $\mathrm{\psi }_{\dot{\alpha}}$
with label $\dot{\alpha}=\pm $) such that the gauge transformation reads as $%
\mathrm{\psi }_{\dot{\alpha}}=U_{\dot{\alpha}}^{\alpha }\Psi _{\alpha }$;
that is $\mathrm{\psi }_{\pm }=U_{\pm \downarrow }\Psi _{\uparrow }+U_{\pm
\uparrow }\Psi _{\downarrow }$. This local rotation of the electronic spin
wave induces a non abelian gauge potential with components $\mathcal{A}_{\mu
}=-iU\partial _{\mu }U^{\dagger }$ mediating the interaction between the
electron and the magnetic texture. Indeed, putting the unitary change into $%
\mathcal{L}_{e}\left[ \Psi ,\mathbf{n}\right] $, we end up with an
equivalent Lagrangian density; but now with new field variables as follows%
\begin{equation}
\mathcal{L}_{e}\left[ \mathbf{\psi },\mathcal{A}_{\mu }\right] =\hbar
\mathbf{\psi }^{\dagger }\left( i\partial _{0}-A_{0}^{a}\sigma _{a}\right)
\mathbf{\psi }-\mathbf{\psi }^{\dagger }\left( \frac{\left( \mathbf{P}+\hbar
\mathbf{A}^{a}\sigma _{a}\right) ^{2}}{2m}-J_{H}\sigma ^{z}\right) \mathbf{%
\psi }  \label{el}
\end{equation}%
Here, the vector potential matrix $\mathcal{A}_{\mu }$ is valued in the $%
SU\left( 2\right) $ Lie algebra generated by the Pauli matrices $\sigma ^{a}$%
; so it can be expanded as $A_{\mu }^{x}\sigma ^{x}+A_{\mu }^{y}\sigma
^{y}+A_{\mu }^{z}\sigma ^{z}$ with components $A_{\mu }^{a}=\frac{1}{2}%
Tr\left( \sigma ^{a}\mathcal{A}_{\mu }\right) $. Notice that in going from
the old $\mathcal{L}_{e}\left[ \Psi ,\mathbf{n}\right] $ to the new $%
\mathcal{\tilde{L}}_{e}\left[ \mathbf{\psi },\mathcal{A}_{\mu }\right] $,
the spin texture $\mathbf{n}$ has disappeared; but not completely as it is
manifested by an emergent non abelian gauge potential $\mathcal{A}_{\mu }$;
so everything is as if we have an electron interacting with an external
field $\mathcal{A}_{\mu }$. To get the explicit relation between the gauge
potential and the magnetisation, we use the isomorphism $SU\left( 2\right)
\sim \mathbb{S}^{3}$ and the Hopf fibration $\mathbb{S}^{1}\times \mathbb{S}%
^{2}$ to write the unitary matrix $U$ as follows%
\begin{equation}
U=e^{i\gamma }\left(
\begin{array}{cc}
\cos \frac{\theta }{2} & e^{-i\varphi }\sin \frac{\theta }{2} \\
e^{+i\varphi }\sin \frac{\theta }{2} & -\cos \frac{\theta }{2}%
\end{array}%
\right) \quad ,\quad \mathcal{A}_{\mu }=\left(
\begin{array}{cc}
\mathfrak{Z}_{\mu } & W_{\mu }^{-} \\
W_{\mu }^{+} & -\mathfrak{Z}_{\mu }%
\end{array}%
\right)  \label{ua}
\end{equation}%
where the factor $e^{i\gamma }$ describes $\mathbb{S}^{1}$ and where, for
later use, we have set $W_{\mu }^{\pm }=A_{\mu }^{1}\pm iA_{\mu }^{2}$ and $%
\mathfrak{Z}_{\mu }=A_{\mu }^{3}$. So, a specific realisation of the gauge
transformation is given by fixing $\gamma =cst$ ( say $\gamma =0$)$;$ it
corresponds to restricting $\mathbb{S}^{3}$ down to $\mathbb{S}^{2}$ and SU$%
\left( 2\right) $ reduces down to SU$\left( 2\right) $/U$\left( 1\right) $.
In this parametrisation, we can also express the unitary matrix U like $%
\mathbf{m.\sigma }$ with magnetic vector $\mathbf{m}=\left( \sin \frac{%
\theta }{2}\cos \varphi ,\sin \frac{\theta }{2}\sin \varphi ,\cos \frac{%
\theta }{2}\right) $ obeying the property $\mathbf{m}^{2}=1;$ the same
constraint as before. By putting back into $U\mathbf{\sigma }.\mathbf{n}%
U^{\dagger }$, and using some algebraic relations like $\varepsilon
_{abd}\varepsilon _{dce}=\delta _{ac}\delta _{be}-\delta _{bc}\delta _{ae},$
we obtain $\left[ 2\left( \mathbf{m}.\mathbf{n}\right) \mathbf{m}-\mathbf{n}%
\right] .\mathbf{\sigma }$. Then, substituting $\mathbf{n}$ by its
expression $\left( \sin \theta \cos \varphi ,\sin \theta \sin \varphi ,\cos
\theta \right) $, we end up with the desired direction $\sigma ^{z}$
appearing in eq(\ref{el}). On the other hand, by putting $U=\mathbf{m.\sigma
}$ back into $-iU\partial _{\mu }U^{\dagger }$, we obtain an explicit
relation between the gauge potential and the magnetic texture namely $A_{\mu
}^{a}=\varepsilon ^{abc}m_{b}\partial _{\mu }m_{c}$. From this expression,
we learn the entries of the potential matrix $\mathcal{A}_{\mu }$ of eq(\ref%
{ua}); the relation with the texture $\mathbf{n}$ is given in what follows
seen that $\mathbf{m}\left( \theta \right) \mathbf{=n}(\theta /2)$.

\subsubsection{Large Hund coupling limit}

We start by noticing that the non abelian gauge potential $A_{\mu }^{a}$
obtained above can be expressed in a condensed form like $\varepsilon
^{abc}m_{a}\partial _{\mu }m_{b}$ (for short $\mathbf{m}\wedge \mathbf{%
\partial }_{\mu }\mathbf{m})$; so it is normal to $\mathbf{m}$; and then it
can be expanded as follows%
\begin{equation}
A_{\mu }^{a}=\frac{1}{2}\mathrm{e}^{a}\partial _{\mu }\theta -\mathrm{f}%
^{a}\sin \frac{\theta }{2}\partial _{\mu }\varphi
\end{equation}%
where we have used the local basis vectors $\mathbf{m}(\theta \mathbf{),e}%
(\theta )$ and $\mathbf{f}(\theta )$. This is an orthogonal triad which turn
out to be intimately related with the triad vectors given by eq(\ref{1});
the relationships read respectively like $\mathbf{n}(\theta /2)\mathbf{,u}%
(\theta /2)$ and $\mathbf{v}(\theta /2)$ involving $\theta /2$ angle instead
of $\theta $. Substituting these basis vectors by their angular values, we
obtain%
\begin{equation}
\left(
\begin{array}{c}
A_{\mu }^{1} \\
A_{\mu }^{2} \\
A_{\mu }^{3}%
\end{array}%
\right) =\frac{1}{2}\left(
\begin{array}{c}
-\sin \varphi \\
\cos \varphi \\
0%
\end{array}%
\right) \partial _{\mu }\theta -\frac{1}{2}\left(
\begin{array}{c}
\sin \theta \cos \varphi \\
\sin \theta \sin \varphi \\
\cos \theta -1%
\end{array}%
\right) \partial _{\mu }\varphi
\end{equation}%
from which we learn that the two first components combine in a complex gauge
field $W_{\mu }^{\pm }=A_{\mu }^{1}\pm iA_{\mu }^{2}$ which is equal to $%
\frac{i}{2}e^{i\varphi }\mathbf{w}^{\pm }\partial _{\mu }\mathbf{n}$ with $%
\mathbf{w^{\pm }=u}\pm i\mathbf{v}$; and the third component $A_{\mu }^{3}$
has the remarkable form $\frac{1}{2}\left( 1-\cos \theta \right) \partial
_{\mu }\varphi $ whose structure recalls the geometric Berry term (\ref{bl}%
). Below, we set $A_{\mu }^{3}=\mathfrak{Z}_{\mu }$ as in eq(\ref{ua}); it
contains the temporal component $\mathfrak{Z}_{0}$ and the three spatial
ones $\mathfrak{Z}_{i}$ --- denoted in section 2 respectively as $a_{0}$ and
$a_{i}$---.\newline
In the large Hund coupling ($J_{H}>>1$), the spin of the electron is quasi-
aligned with the magnetisation $\mathbf{n}$; so the electronic dynamics is
mainly described by the chiral wave function $\left( \mathrm{\psi }%
_{+},0\right) $ denoted below as $\mathbf{\chi }=\left( \mathrm{\chi }%
,0\right) $. Thus, the effective properties of the interaction between the
electron and the skyrmion can be obtained by restricting the above relations
to the polarised electronic spin wave $\mathbf{\chi }$. By setting $\mathrm{%
\psi }_{-}=0$ into eq(\ref{el}) and using $\mathbf{\chi }^{\dagger }\sigma
^{x}\mathbf{\chi }=\mathbf{\chi }^{\dagger }\sigma ^{y}\mathbf{\chi }=0$ and
$\mathbf{\chi }^{\dagger }\sigma ^{z}\mathbf{\chi }=\mathrm{\bar{\chi}\chi }$
\ as well as replacing $\left( A_{\mu }^{x}\sigma _{x}\right) ^{2}+\left(
A_{\mu }^{y}\sigma _{y}\right) ^{2}$ by $\frac{1}{4}\left( \partial _{\mu }%
\mathbf{n}\right) ^{2}$, the Lagrangian (\ref{el}) reduces to the polarised $%
\mathcal{L}_{e}^{\text{{\small (pol)}}}=\mathcal{L}_{e}\left[ \mathrm{\chi },%
\mathbf{n},\mathcal{Z}_{\mu }\right] $ given by%
\begin{equation}
\mathcal{L}_{e}^{\text{{\small (pol)}}}=\hbar \mathbf{\chi }^{\dagger
}\left( i\partial _{0}-\mathfrak{Z}_{0}\sigma ^{z}\right) \mathbf{\chi }-%
\mathbf{\chi }^{\dagger }\left( \frac{\left( P_{i}+\hbar \mathfrak{Z}%
_{i}^{a}\sigma _{a}\right) ^{2}}{2m}+\frac{\hbar ^{2}}{8m}\left( \partial
_{\mu }\mathbf{n}\right) ^{2}-J_{H}\sigma ^{z}\right) \mathbf{\chi }
\label{le}
\end{equation}%
where $\left( \mathfrak{Z}_{0},\mathfrak{Z}_{i}\right) $ define the four
components of the emergent abelian gauge prepotential $\mathfrak{Z}_{\mu }$
associated with the Pauli matrix $\sigma ^{z}$; their explicit expressions
are given by $\mathfrak{Z}_{0}=\frac{1}{2}\left( 1-\cos \theta \right) \dot{%
\varphi}$ and $\mathfrak{Z}_{i}=\frac{1}{2}\left( 1-\cos \theta \right)
\partial _{i}\varphi $; their variation with respect to the magnetic texture
are related to the magnetisation field like $\frac{\delta \mathfrak{Z}_{\mu }%
}{\delta \mathbf{n}}=\frac{1}{2}\partial _{\mu }\mathbf{n\wedge n}$.

\subsection{Skyrmion with spin transfer torque}

Here, we investigate the full dynamics of the electron/skyrmion system $%
\left \{ e^{-},\mathbf{n}\right \} $ described by the Lagrangian density $%
\mathcal{L}_{tot}$ containing the parts $\mathcal{L}_{\mathbf{n}}+\mathcal{L}%
_{e-\mathbf{n}};$ the electronic Lagrangian $\mathcal{L}_{e-\mathbf{n}}$ is
given by eq(\ref{el}). The Lagrangian $\mathcal{L}_{\mathbf{n}}$, describing
the skyrmion dynamics, is as in eqs(\ref{lb}-\ref{bl}) namely $-\hbar S%
\mathfrak{Z}_{0}-\mathcal{H}_{\mathbf{n}}$ with $\mathfrak{Z}_{0}=\frac{1}{2}%
\left( 1-\cos \theta \right) \dot{\varphi}.$ By setting $\mathcal{\tilde{H}}%
_{\mathbf{n}}=\mathcal{H}_{\mathbf{n}}+\frac{\hbar ^{2}}{8m}\left( \partial
_{\mu }\mathbf{n}\right) ^{2}\mathbf{\psi }^{\dagger }\mathbf{\psi }$, the
full Lagrangian density $\mathcal{L}_{tot}$ with can be then presented like $%
\mathcal{\tilde{L}}\left[ \mathbf{\psi ,}\mathfrak{Z}_{\mu }\right] -\tilde{H%
}_{\mathbf{n}}$ like
\begin{equation}
\mathcal{\tilde{L}}\left[ \mathbf{\psi ,}\mathfrak{Z}_{\mu }\right] =-\hbar S%
\mathfrak{Z}_{0}+\hbar \mathbf{\psi }^{\dagger }\left( i\frac{\partial }{%
\partial t}-\mathfrak{Z}_{0}\sigma ^{z}\right) \mathbf{\psi }-\mathbf{\psi }%
^{\dagger }\left( \frac{\left( P_{i}+\hbar \mathfrak{Z}_{i}\sigma
^{z}\right) ^{2}}{2m}\right) \mathbf{\psi }  \label{tl}
\end{equation}%
with $\left( P_{i}+\hbar \mathfrak{Z}_{i}\sigma ^{z}\right) ^{2}$ expanding
as $P_{i}^{2}+\hbar _{i}^{2}\mathfrak{Z}^{2}+\hbar \left( P_{i}\mathfrak{Z}%
^{i}+\mathfrak{Z}^{i}P_{i}\right) \sigma ^{z}$. The equations of motion of $%
\mathbf{\psi }$ and $\mathbf{n}$ are obtained as usual by computing the
extremisation of this Lagrangian density with respect to the corresponding
field variables. In general, we have $\delta \mathcal{L}_{tot}=\left( \delta
\mathcal{L}_{tot}/\delta \mathbf{n}\right) .\delta \mathbf{n}+\left( \delta
\mathcal{L}_{tot}/\delta \mathbf{\psi }\right) .\delta \mathbf{\psi }+hc$
which vanishes for $\delta \mathcal{L}_{tot}/\delta \mathbf{n}=0$ and $%
\delta \mathcal{L}_{tot}/\delta \mathbf{\psi }^{\dagger }=0$.

\subsubsection{Modified Landau- Lifshitz equation}

Regarding the spin texture $\mathbf{n}$, the associated field equation of
motion is given by $\delta \mathcal{L}_{tot}/\delta \mathbf{n}=0$; the
contributions to this equation of motion come from the variations $\mathcal{%
\tilde{L}}$ and $\mathcal{\tilde{H}}_{\mathbf{n}}$ with respect to $\delta
\mathbf{n}$ namely%
\begin{equation}
\frac{\delta \mathcal{\tilde{H}}}{\delta \mathbf{n}}-\frac{\delta \mathcal{%
\tilde{L}}}{\delta \mathfrak{Z}_{\mu }}\frac{\delta \mathfrak{Z}_{\mu }}{%
\delta \mathbf{n}}=0  \label{hll}
\end{equation}
The variation $\frac{\delta \mathcal{H}_{\mathbf{n}}}{\delta \mathbf{n}}$
depends on the structure of the skyrmion Hamiltonian density $\mathcal{%
\tilde{H}}$; its contribution to the equation of motion can be presented
like $\lambda \partial ^{\mu }\partial _{\mu }\mathbf{n}=\mathbf{F}$ with
some factor $\lambda $. However, the variation $\frac{\delta \mathcal{L}}{%
\delta \mathfrak{Z}_{\mu }}\frac{\delta \mathfrak{Z}_{\mu }}{\delta \mathbf{n%
}}$ describes skyrmion-electron interaction; and can be done explicitly into
two steps; the first step concerns the calculation of the time like
component $\frac{\delta \mathcal{L}}{\delta \mathfrak{Z}_{0}}\frac{\delta
\mathfrak{Z}_{0}}{\delta \mathbf{n}}$; it gives $-\frac{\hbar }{2}\left[ 2S+%
\mathbf{\psi }^{\dagger }\sigma ^{z}\mathbf{\psi }\right] (\mathbf{\dot{n}%
\wedge n});$ it is normal to $\mathbf{n}$ and to velocity $\mathbf{\dot{n}}$
and involves the eletron spin density $\varrho _{e}^{z}=\mathbf{\psi }%
^{\dagger }\sigma ^{z}\mathbf{\psi }$. \newline
The second step deals with the calculation of the space like component $-%
\frac{\delta \mathcal{L}}{\delta \mathfrak{Z}_{i}}\frac{\delta \mathfrak{Z}%
_{i}}{\delta \mathbf{n}}$; the factor $\frac{\delta \mathcal{L}}{\delta
\mathfrak{Z}_{i}}$ gives $-\hbar \mathcal{J}^{i}$ with a 3-component current
vector density reading as follows
\begin{equation}
\mathcal{J}_{i}=\frac{1}{2m}\left( \mathbf{\psi }^{\dagger }\sigma ^{z}P_{i}%
\mathbf{\psi }-P_{i}\mathbf{\psi }^{\dagger }\sigma ^{z}\mathbf{\psi }%
\right) +\frac{\hbar }{m}(\mathbf{\psi }^{\dagger }\mathbf{\psi )}\mathfrak{Z%
}_{i}  \label{j}
\end{equation}
This vector two remarkable properties: $\left( 1\right) $ it is given by the
sum of two contributions as it it reads like $\mathcal{J}_{i}^{\left( +%
\mathbf{n}\right) }+\mathcal{J}_{i}^{\left( -\mathbf{n}\right) }$ with%
\begin{equation}
\begin{tabular}{lll}
$\mathcal{J}_{l}^{\left( +\mathbf{n}\right) }$ & $=$ & $\frac{\hbar }{m}(%
\mathrm{\bar{\psi}}_{+\mathbf{n}}\mathrm{\psi }_{+\mathbf{n}}\mathbf{)}%
\mathfrak{Z}_{l}+\frac{1}{2m}\left( \mathrm{\bar{\psi}}_{+\mathbf{n}}\frac{%
\hbar }{i}\partial _{l}\mathrm{\psi }_{+\mathbf{n}}-\frac{\hbar }{i}\partial
_{l}\mathrm{\bar{\psi}}_{+\mathbf{n}}\mathrm{\psi }_{+\mathbf{n}}\right) $
\\
$\mathcal{J}_{l}^{\left( -\mathbf{n}\right) }$ & $=$ & $\frac{\hbar }{m}(%
\mathrm{\bar{\psi}}_{-\mathbf{n}}\mathrm{\psi }_{-\mathbf{n}}\mathbf{)}%
\mathfrak{Z}_{l}-\frac{\hbar }{2im}\left( \mathrm{\bar{\psi}}_{-\mathbf{n}}%
\frac{\hbar }{i}\partial _{l}\mathrm{\psi }_{-\mathbf{n}}-\frac{\hbar }{i}%
\partial _{l}\mathrm{\bar{\psi}}_{-\mathbf{n}}\mathrm{\psi }_{-\mathbf{n}%
}\right) $%
\end{tabular}%
\end{equation}%
These vectors are respectively interpreted as two spin polarised currents;
the $\mathcal{J}_{i}^{\left( +\mathbf{n}\right) }$ is associated with the $%
\mathrm{\psi }_{+\mathbf{n}}$ wave function as it points in the same
direction as $\mathbf{n}$; the $\mathcal{J}_{i}^{\left( -\mathbf{n}\right) }$
is however associated with $\mathrm{\psi }_{-\mathbf{n}}$ pointing in the
opposite direction of $\mathbf{n}$. $\left( 2\right) $ Each one of the two $%
\mathcal{J}^{\left( +\mathbf{n}\right) }$ and $\mathcal{J}^{\left( -\mathbf{n%
}\right) }$ are in turn given by the sum of two contributions as they can be
respectively split like $\frac{\hbar }{m}(\mathrm{\bar{\psi}}_{+\mathbf{n}}%
\mathrm{\psi }_{+\mathbf{n}}\mathbf{)}\mathfrak{Z}+\mathbf{j}_{\mathrm{\psi }%
_{+\mathbf{n}}}$ and $\frac{\hbar }{m}(\mathrm{\bar{\psi}}_{-\mathbf{n}}%
\mathrm{\psi }_{-\mathbf{n}}\mathbf{)}\mathfrak{Z}+\mathbf{j}_{\mathrm{\psi }%
_{-\mathbf{n}}}$ with vector density $\mathbf{j}_{\mathrm{\psi }}$ standing
for the usual current vector $\mathbf{j}_{\mathrm{\psi }}=\frac{1}{2m}%
\mathrm{\bar{\psi}}\overleftrightarrow{\mathbf{P}}\mathrm{\psi }$. The
contribution $\frac{\hbar }{m}(\mathrm{\bar{\psi}\psi }\mathbf{)}\mathfrak{Z}
$ is proportional to the emergent gauge field $\mathfrak{Z}$; it defines a
spin torque transfert to the vector current density $\mathcal{J}_{i}$.%
\newline
Regarding the factor $\frac{\delta \mathfrak{Z}_{i}}{\delta \mathbf{n}}$, it
gives $\frac{1}{2}(\partial _{i}\mathbf{n\wedge n})$; by substituting, the
total contribution of $\frac{\delta \mathcal{L}}{\delta \mathfrak{Z}_{i}}%
\frac{\delta \mathfrak{Z}_{i}}{\delta \mathbf{n}}$ leads to $-\frac{\hbar }{2%
}(\mathcal{J}^{i}\partial _{i}\mathbf{n})\mathbf{\wedge n}$ that reads in a
condensed form like $-\frac{\hbar }{2}(\mathcal{J}.\nabla \mathbf{n})\mathbf{%
\wedge n}$. Putting back into eq(\ref{hll}), we end up with the following
modified LL equation%
\begin{equation}
-\frac{\hbar }{2}\left[ 2S+\mathbf{\psi }^{\dagger }\sigma ^{z}\mathbf{\psi }%
\right] (\mathbf{\dot{n}\wedge n})+\frac{\hbar }{2}(\mathcal{J}.\nabla
\mathbf{n})\mathbf{\wedge n}-\frac{\delta H_{\mathbf{n}}}{\delta \mathbf{n}}%
=0  \label{eqn}
\end{equation}%
To compare this equation with the usual LL equation ($\hbar S\mathbf{\dot{n}}%
=\frac{\delta H_{\mathbf{n}}}{\delta \mathbf{n}}\wedge \mathbf{n)}$ in
absence of Hund coupling (which corresponds to putting $\mathbf{\psi }$ to
zero), we multiply eq(\ref{eqn}) by $\wedge \mathbf{n}$ in order to bring it
to a comparable relation with LL equation. By setting $\varrho _{e}^{z}=%
\mathbf{\psi }^{\dagger }\sigma ^{z}\mathbf{\psi }$, describing the
electronic spin density $|\mathrm{\psi }_{+\mathbf{n}}|^{2}-|\mathrm{\psi }%
_{-\mathbf{n}}|^{2}$; we find%
\begin{equation}
-\hbar \left[ \frac{S}{\text{\texttt{a}}^{-d}}+\frac{\varrho _{e}^{z}}{2}%
\right] \mathbf{\dot{n}}=\frac{\delta H_{\mathbf{n}}}{\delta \mathbf{n}}%
\wedge \mathbf{n}-\hbar \left[ \left( \mathcal{J}.\nabla \right) \mathbf{n}%
\right]
\end{equation}%
where, due to $\mathbf{n}^{2}=1$, the space gradient $\mathcal{J}.\nabla
\mathbf{n}$ is normal to $\mathbf{n}$; and so it can be set as $\mathbf{%
\Omega }^{\left( e\right) }\wedge \mathbf{n}$ with $\mathbf{\Omega }^{\left(
e\right) }=\mathcal{J}^{i}\mathbf{\omega }_{i}^{\left( e\right) }$. The
above equation is a modified LL equation; it describes the dynamics of the
spin texture interacting with electrons through Hund coupling. Notice that
for $\mathbf{\psi }$ $\rightarrow 0$, this equation reduces to $\hbar \frac{S%
}{\text{\texttt{a}}^{-d}}\mathbf{\dot{n}}=\mathbf{\omega }^{\left( n\right)
}\wedge \mathbf{n}$ showing that the vector $\mathbf{n}$ rotates with $%
\mathbf{\omega }^{\left( n\right) }=-\frac{\delta H_{\mathbf{n}}}{\delta
\mathbf{n}}$. By turning on $\mathbf{\psi }$, we have $\mathbf{\dot{n}}\sim (%
\mathbf{\omega }^{\left( n\right) }+\mathbf{\Omega }^{\left( e\right)
})\wedge \mathbf{n}$ indicating that the LL rotation is drifted by $\mathbf{%
\Omega }^{\left( e\right) }$ coming from two sources: $\left( i\right) $ the
term $\hbar \left[ \left( \mathcal{J}.\nabla \right) \mathbf{n}\right] $
which deforms LL vector $\mathbf{\omega }^{\left( n\right) }$ drifted by the
$\mathbf{n}\wedge \left( \mathcal{J}.\nabla \mathbf{n}\right) $; and $\left(
ii\right) $ the electronic spin density $\varrho _{e}^{z}=\frac{N_{e}}{\text{%
\texttt{a}}^{-d}}$; this term adds to the density $\frac{S}{\text{\texttt{a}}%
^{-d}}$ of the magnetic texture per unit volume; it involves the number $%
N_{e}=N_{e}^{+\mathbf{n}}-N_{e}^{-\mathbf{n}}$ with $N_{e}^{\pm \mathbf{n}}$
standing for the filling factor of polarized conduction electrons. Moreover,
if assuming $\mathbf{n}\left( t,\mathbf{r}\right) =\mathbf{n}\left( \mathbf{r%
}-\mathbf{V}_{s}t\right) $ with a uniform $\mathbf{V}_{s}$, then the drift
velocity $\dot{n}^{a}=-\left( \partial _{i}n^{a}\right) V_{s}^{i}$ and $%
\left( J_{e}^{i}\partial _{i}\right) n^{a}=J_{e}^{a}$. Putting back into the
modified LLG equation, we end up with the following relation between the $%
\mathbf{V}_{s}$ and $\mathbf{v}_{e}$ velocities $(S+\frac{n_{e}}{2}%
)v_{s}^{a}=n_{e}v_{e}^{a}$ where we have set $\left( \partial
_{i}n^{a}\right) V_{s}^{i}=v_{s}^{a}$ and $J_{e}^{a}=n_{e}v_{e}^{a}$.

\subsubsection{Rigid skyrmion under spin transfer torque}

Here, we investigate the dynamics of a 2D rigid skyrmion [$\mathbf{n}=%
\mathbf{n}\left( \mathbf{r}-\mathbf{R}\right) ]$ under a spin transfer
torque (STT) induced by itinerant electrons. For that, we apply the method,
used in sub-subsection 5.1.2 to derive $L_{s}$ from the computation space
integral of $\int d^{2}\mathbf{r}\mathcal{L}_{s}$ and eqs(\ref{lh}). To
begin, recall that in absence\ of the STT effect, the Lagrangian $L_{s}$ of
the 2D skyrmion's point- particle, with position $\mathbf{R}=(X,Y)$ and
velocity $\mathbf{\dot{R}}=(\dot{X},\dot{Y}),$ is given by $\frac{M_{s}}{2}%
\mathbf{\dot{R}}^{2}-\frac{G}{2}\mathbf{z.(R}\wedge \mathbf{\dot{R}})-V(%
\mathbf{R})$ with effective scalar energy potential $V(\mathbf{R})=\int d^{2}%
\mathbf{r}\mathcal{H}\left( \mathbf{r,R}\right) $ and a constant $G=\frac{%
4\pi \hbar }{\text{\texttt{a}}^{2}}q_{s}S.$ Under STT induced by Hund
coupling, the Lagrangian $L_{s}$ gets deformed into $\mathtt{\tilde{L}}%
_{s}=L_{s}+\Delta L_{s}$, that is%
\begin{equation}
\mathtt{\tilde{L}}_{s}=\frac{M_{s}}{2}\mathbf{\dot{R}}^{2}-\frac{G}{2}%
\mathbf{z.(R}\wedge \mathbf{\dot{R}})-V(\mathbf{R})+\Delta L_{s}  \label{de}
\end{equation}%
To determine $\Delta L_{s}$, we start from $\mathtt{\tilde{L}}_{s}=\int d^{2}%
\mathbf{r}\mathcal{\tilde{L}}_{tot}$ with Lagrangian density as $\mathcal{%
\tilde{L}}_{tot}=\mathcal{\tilde{L}}-\mathcal{\tilde{H}}_{\mathbf{n}}$ with $%
\mathcal{\tilde{L}}$ given by eq(\ref{tl}). For convenience, we set $%
\mathcal{\tilde{L}}=-\hbar S\mathfrak{Z}_{0}+\mathcal{\tilde{L}}_{e-\mathbf{n%
}}$ and set%
\begin{equation}
\mathcal{\tilde{L}}_{e-\mathbf{n}}=\hbar \mathbf{\psi }^{\dagger }\left( i%
\frac{\partial }{\partial t}-\mathfrak{Z}_{0}\sigma ^{z}\right) \mathbf{\psi
}-\mathbf{\psi }^{\dagger }\left( \frac{\left( P_{i}+\hbar \mathfrak{Z}%
_{i}\sigma ^{z}\right) ^{2}}{2m}\right) \mathbf{\psi }  \label{sen}
\end{equation}%
The deviation $\Delta L_{s}$ with respect to $L_{s}$ in (\ref{de}) comes
from those terms in eq(\ref{sen}). Notice that this expression involves the
wave function $\mathbf{\psi }$ coupled to the emergent gauge potential field
$\mathfrak{Z}_{\mu }=\left( \mathfrak{Z}_{0},\mathfrak{Z}_{i}\right) $; that
is $-\hbar \int d^{2}\mathbf{r\psi }^{\dagger }\sigma ^{z}\mathbf{\psi }%
\mathfrak{Z}_{0}$ and $-\frac{1}{2m}\int d^{2}\mathbf{r\psi }^{\dagger
}[\left( P_{i}+\hbar \mathfrak{Z}_{i}\sigma ^{z}\right) ^{2}]\mathbf{\psi }$%
. Thus, to obtain $\Delta L_{s}$, we first calculate the variation $\frac{%
\delta \left( \Delta L_{s}\right) }{\delta \mathfrak{Z}_{\mu }}\delta
\mathfrak{Z}_{\mu }$ and put $\delta \mathfrak{Z}_{\mu }=\frac{\delta
\mathfrak{Z}_{\mu }}{\delta \mathbf{R}}.\delta \mathbf{R}$. Once, we have
the explicit expression of this variation, we turn backward to deduce the
value of $\Delta L_{s}$. To that purpose, we proceed in two steps as
follows: $\left( i\right) $ We calculate the temporal contribution $\frac{%
\delta \left( \Delta L_{s}\right) }{\delta \mathfrak{Z}_{0}}\frac{\delta
\mathfrak{Z}_{0}}{\delta \mathbf{R}}.\delta \mathbf{R}$; and $\left(
ii\right) $ we compute the spatial $\frac{\delta \left( \Delta L_{s}\right)
}{\delta \mathfrak{Z}_{i}}\frac{\delta \mathfrak{Z}_{i}}{\delta \mathbf{R}}%
.\delta \mathbf{R.}$ Using the variation $\delta \mathfrak{Z}_{0}=\frac{1}{2}%
\delta \mathbf{n.}\left( \mathbf{n\wedge \partial }_{j}\mathbf{n}\right)
\dot{X}^{j}$, the contribution of the first term can be put as follows
\begin{equation}
\frac{\delta \left( \Delta L_{s}\right) }{\delta \mathfrak{Z}_{0}}\frac{%
\delta \mathfrak{Z}_{0}}{\delta X^{l}}\delta X^{l}\mathbf{=}-\frac{\hbar }{2}%
\boldsymbol{J}_{0}^{z}\epsilon _{zij}\left[ \dot{X}^{i}\delta X^{j}\right]
\end{equation}%
where we have set $\varrho ^{z}=\mathbf{\psi }^{\dagger }\sigma ^{z}\mathbf{%
\psi }$ and $\boldsymbol{J}_{0}^{z}=\int d^{2}\mathbf{r}\frac{\varrho ^{z}}{2%
}\epsilon ^{zkl}\mathbf{n.}\left( \partial _{k}\mathbf{n\wedge }\partial _{l}%
\mathbf{n}\right) .$ Notice that the right hand side in above relation can
be also put into the form $\frac{\hbar }{2}\boldsymbol{J}_{0}^{z}\epsilon
_{zij}\left[ \delta \dot{X}^{i}X^{j}\right] -\delta \left[ \frac{\hbar }{2}%
\epsilon _{zij}\boldsymbol{J}_{0}^{z}\dot{X}^{i}X^{j}\right] $ indicating
that $\Delta L_{s}$ must contain the term $\frac{\hbar }{2}\epsilon _{zij}%
\boldsymbol{J}_{0}^{z}\dot{X}^{i}X^{j}$ which reads as well like $\frac{%
\hbar }{2}J_{0}\mathbf{z.}\left( \mathbf{\dot{R}}\wedge \mathbf{R}\right) $.
Regarding the spatial part $\frac{\delta \left( \Delta L_{s}\right) }{\delta
\mathfrak{Z}_{i}}.\frac{\delta \mathfrak{Z}_{i}}{\delta X^{l}}\delta X^{l}$,
we have quite similar calculations allowing to put it in the following form%
\begin{equation}
\frac{\delta \left( \Delta L_{s}\right) }{\delta \mathfrak{Z}_{i}}.\frac{%
\delta \mathfrak{Z}_{i}}{\delta X^{l}}\delta X^{l}=-\hbar \varepsilon _{zij}%
\boldsymbol{J}^{zi}\delta X^{j}
\end{equation}%
where we have set $\boldsymbol{J}^{zi}\left( t\right) =\int d^{2}\mathbf{r}%
\mathcal{J}^{zi}\left( t,\mathbf{r}\right) $ with $\mathcal{J}^{zi}\left( t,%
\mathbf{r}\right) $ given by eq(\ref{j}). Here also notice that the right
hand of above equation can be put as well like $\delta \left[ -\hbar
\varepsilon _{zij}\boldsymbol{J}^{zi}X^{j}\right] $ indicating that $\Delta
L_{s}$ contains in addition to $\frac{\hbar }{2}J_{0}\mathbf{z.}\left(
\mathbf{\dot{R}}\wedge \mathbf{R}\right) $, the term $-\hbar \varepsilon
_{zij}\boldsymbol{J}^{zi}X^{j}$ which reads also as $-\hbar \mathbf{z.}%
\boldsymbol{J}\wedge \mathbf{R}$ with two component vector $\boldsymbol{J}%
=\left( \boldsymbol{J}^{zx},\boldsymbol{J}^{zy}\right) $. Thus, we have the
following modified skyrmion equation%
\begin{equation}
\mathtt{\tilde{L}}_{s}=\frac{M_{s}}{2}\mathbf{\dot{R}}^{2}-\frac{1}{2}\left(
G+\hbar J_{0}\right) \mathbf{z.}(\mathbf{\dot{R}}\wedge \mathbf{R)}+\hbar
\mathbf{z.}\left( \boldsymbol{J}\wedge \mathbf{R}\right) -V(\mathbf{R)}
\end{equation}%
from which we determine the modified equation of motion of the rigid
skyrmion in presence of spin transfer torque.

\section{Comments and perspectives}

In this bookchapter, we have studied the basic aspects of the solitons
dynamics in various $\left( 1+d\right) $ spacetime dimensions with $d=1,2,3$%
; while emphasizing the analysis of their topological properties and their
interaction with the environment. After having introduced the quantum SU$%
\left( 2\right) $ spins, their coherent vector representation $\mathbf{S}=%
\mathcal{R}\left( \alpha ,\beta ,\gamma \right) \mathbf{S}_{0}$ with $%
\mathbf{S}_{0}$ standing for the highest weight spin state; and their link
with the magnetic moments $\mathbf{\mu \ltimes }S\mathbf{n}$, we have
revisited the time evolution of coherent spin states; and proceeded by
investigating their spatial distribution while focusing on kinks, 2d and 3d
skyrmions. We have also considered the rigid skyrmions dissolved in the
magnetic texture without and with dissipation. Moreover, we explored the
interaction between electrons and skyrmions and analyzed the effect of the
spin transfer torque. In this regard, we have refined the results concerning
the modified LL equation for the rigid skyrmion in connection with emergent
non abelian SU$\left( 2\right) $ gauge fields. It is found that the magnetic
skyrmions, existing in a ferromagnetic (FM) medium, show interesting
behaviors such as emergent electrodynamics \textrm{\cite{A0012}} and
current-driven motion at low current densities \textrm{\cite{A007,A008}}.
Consequently, the attractive properties of ferromagnetic skyrmions make them
promising candidates for high-density and low-power spintronic technology.
Besides, ferromagnetic skyrmions have the potential to encode bits in
low-power magnetic storage devices. Therefore, alternative technology of
forming and controlling skyrmions is necessary for their use in device
engineering. This investigation was performed by using the field theory
method based on coherent spin states described by a constrained spacetime
field captured by $f\left( \mathbf{n}\right) =1$. Such condition supports
the topological symmetry of magnetic solitons which is found to be
characterised by integral topological charges $Q$ that are interpreted in
terms of magnetic skyrmions and antiskyrmion; these topological states can
be imagined as (winding) quasiparticle excitations with $Q>0$ and $Q<0$
respectively.

Regarding these two skyrmionic configurations, it is interesting to notice
that, unlike magnetic skyrmions, the missing rotational symmetry of
antiskyrmions leads to anisotropic DMI, which is highly relevant for
racetrack applications. It follows that antiskyrmions exist in certain
Heusler materials having a particular type of DMI, including MnPtPdSn \cite%
{zz13} and MnRhIrSn \textrm{\cite{rrr143}}. It is then deduced that
stabilized antiskyrmions can be observed in materials exhibiting D$_{2d}$
symmetry such as layered systems with heavy metal atoms. Furthermore, the
antiskyrmion show some interesting features, namely long lifetimes at room
temperature and a parallel motion to the applied current \cite{RRR144}.
Thus, antiskyrmions are easy to detect using conventional experimental
techniques and can be considered as the carriers of information in racetrack
devices.

To lift the limitations associated with ferromagnetic\textbf{\ }skyrmions
for low-power spintronic devices, recent trends combine multiple
subparticles in different magnetic surroundings. Stable room-temperature
antiferromagnetic skyrmions in synthetic Pt/Co/Ru antiferromagnets result
from the combination of two FM nano-objects coupled antiferromagnetically
\textrm{\cite{xx11}}. Compared to their ferromagnetic analogs,
antiferromagnetic skyrmions exhibit different dynamics and are driven with
several kilometers per second by currents. Coupling two subsystems with
mutually reversed spins, gives rise to ferrimagnetic skyrmions as detected
in GdFeCo films using scanning transmission X-ray microscopy \textrm{\cite%
{zz032}}. At ambient temperature, these skyrmions move at a speed of $50\,m/s
$ with a reduced skyrmion Hall angle of\thinspace $\ 20%
{{}^\circ}%
$. Characterized by uncompensated magnetization, the vanishing angular
momentum line can be utilized as a self-focusing racetrack for skyrmions.
Another technologically promising object is generated by the coexistence of
skyrmions and antiskyrmions in materials with D$_{2d}$ symmetry. The
resulting spin textures constitute information bits 0' and `1' generalizing
the concept of racetrack device. Insensitive to the repulsive interaction
between the two distinct nano-objects, such emergent devices are promising
solution for racetrack storage applications.

\section*{Acknowledgement}

L. B. Drissi would like to acknowledge "Acad\'{e}mie Hassan II des Sciences
et Techniques-Morocco". She also acknowledges the Alexander von Humboldt
Foundation for financial support via the George Forster Research Fellowship
for experienced scientists (Ref 3.4 - MAR - 1202992).

\end{document}